\documentclass[%
 reprint,
superscriptaddress,
nofootinbib,
 amsmath,amssymb,
 aps,
prx,
floatfix,
longbibliography
]{revtex4-1}

\usepackage{ifpdf}
\usepackage{amsmath} 
\usepackage{amssymb}
\usepackage{amsfonts}
\usepackage{braket}
\usepackage{color}
\usepackage[colorlinks=true,linkcolor=blue,citecolor=blue,breaklinks]{hyperref}
\usepackage{graphicx}
\usepackage{dcolumn}
\usepackage{bm}
\usepackage[normalem]{ulem}

\newcommand{\beq}{\begin{eqnarray}}
\newcommand{\eeq}{\end{eqnarray}}

\newcommand{\la}{\langle}
\newcommand{\ra}{\rangle}

\newcommand{\ie}{i.e.,\ }
\newcommand{\eg}{e.g.,\ }
\newcommand{\trivial}{\mathbf{1}}
\newcommand{\AAtoO}{A\!+\!A\!\rightarrow\!0}
\newcommand{\AAtoA}{A\!+\!A\!\rightarrow\!A}

\definecolor{Zcolour}{rgb}{0.992, 0.588, 0.22}
\definecolor{purple}{rgb}{0.5,0,0.5}
\definecolor{brown}{rgb}{0.6,0.2,0}
\definecolor{dkgreen}{rgb}{0,0.5,0}

\usepackage{mathtools}

\begin{document}


\title{Reaction-diffusion dynamics in a Fibonacci chain:\\ 
Interplay between classical and quantum behavior}
\author{Cheng-Ju Lin}
\affiliation{Perimeter Institute for Theoretical Physics, Waterloo, Ontario, Canada N2L 2Y5}
\author{Liujun Zou}
\thanks{The two authors contributed equally.\\ Corresponding author: zou@perimeterinstitute.ca}
\affiliation{Perimeter Institute for Theoretical Physics, Waterloo, Ontario, Canada N2L 2Y5}


\begin{abstract}

We study the reaction-diffusion dynamics of Fibonacci anyons in a one dimensional lattice.
Due to their non-Abelian nature, besides the position degree of freedom (DOF), these anyons also have a nonlocal internal DOF, which can be characterized by a fusion tree.
We first consider a pure-reaction dynamics associated with the internal DOF, which is of intrinsically quantum origin, with either an ``all-$\tau$" or ``completely random" initial fusion tree.
These two fusion trees are unstable and likely stable steady states for the internal DOF, respectively.
We obtain the decay rate of the anyon number for these two cases exactly.
Still using these two initial fusion trees, we study the full reaction-diffusion dynamics, and find an interesting interplay between classical and quantum behaviors:
These two fusion trees are still respectively unstable and likely stable steady states of the internal DOF, while the dynamics of the position DOF can be mapped to a hybrid classical $A+A\rightarrow 0$ and $A+A\rightarrow A$ reaction-diffuson dynamics, with the relative reaction rates of these two classical dynamics determined by the state of the nonlocal internal DOF.
In particular, the anyon density at late times are given by $\rho(t)=\frac{c}{\sqrt{8\pi}}(Dt)^{-\Delta}$, where $D$ is a non-universal diffusion constant, $\Delta=1/2$ is superuniversal, and $c$ is universal and can be obtained exactly in terms of the fusion tree structure.
Specifically, $c=\frac{2\varphi}{\varphi+1}$ and $c=\frac{2(4\varphi+3)}{5(\varphi+1)}$ for the all-$\tau$ and completely random configuration respectively, where $\varphi=\frac{\sqrt{5}+1}{2}$ is the golden ratio.
We also study the two-point correlation functions.

\end{abstract}

\maketitle
\tableofcontents

\section{Introduction}

Despite the tremendous success in understanding many-body systems at or near thermal equilibrium, it remains a daunting challenge to understand the universal behaviors of systems far from equilibrium.
Relatedly, the available theoretical tools to tackle nonequilibrium dynamics are relatively limited, compared to their counterparts in equilibrium physics. Therefore, it is interesting and useful to obtain exact results regarding nonequilibrium many-body physics, which may shed lights on the more general principles underlying complex many-body dynamics.

Many nonequilibrium systems exhibit a steady state. The nature of the steady state can be viewed as a universal aspect of the underlying dynamics, and there can be phase transitions where the nature of the steady state changes abruptly as the parameters and/or the initial condition of the system are tuned smoothly \cite{Carmichael1980, Werner2004, Capriotti2004, Morrison2008, Eisert2010, Diehl2010, palManybody2010, Bhaseen2011, Kessler2012, Zou2014, Grover2014, luitzManybody2015,serbynCriterion2015a,khemaniCritical2017}.
Just like their equilibrium counterparts, phases defined with respect to the steady states and the transitions between them are often characterized by their correlations at long distances and dynamic responses at low frequencies.

Interestingly, besides the universal low-frequency dynamic responses of a system that already (approximately) reaches a steady state, sometimes a steady state is also associated with a {\it universal late-time dynamics}, \ie the late-time dynamics, which describes how the system approaches a steady state from a given initial state, can display some properties that are in certain sense insensitive to the microscopic details of the system.

Naturally, different steady states can be associated with different universal late-time dynamics. But, in principle, even a {\it single} steady state can also be associated with multiple universality classes of late-time dynamics. Then by slightly varying the parameters and/or initial state of a system (without changing the steady state), one can study the stability of a universal late-time dynamics, \ie under a given type of small perturbation, it is stable (unstable) if the late-time dynamics returns back to (deviates more from) the original universality class. For two stable universal late-time dynamics, one can further study the {\it dynamics transition} between them. These concepts defined for late-time dynamics are natural generalizations of the usual concepts of phases and phase transitions defined for equilibrium and steady states, and it is useful to identify and characterize some concrete examples.

With these general motivations, in this paper, we study the reaction-diffusion dynamics of Fibonacci anyons in one dimension. Reaction-diffusion dynamics is a class of extensively studied nonequilibrium dynamics, and many interesting universal phenomena have been discovered (see, for example, Refs.~\cite{Hinrichsen2000, Tauber2005, Henkel2003} for a review). Its typical setup consists of some particles that can undergo various types of processes, including diffusion, annihilation, coagulation, etc. Such dynamics can be applied to model a wide variety of phenomena, such as relaxations of domain walls, chemical reactions, biological and ecological processes.

As an example, consider many classical particles in one dimension that undergo a type of reaction-diffusion dynamics sometimes referred to as the $A+A\rightarrow 0$ dynamics. These particles can diffuse individually. When two of them are close to each other, at certain annihilation rate, their interactions can turn them into a different particle type, which then immediately escape from the system. Clearly, there is a single steady states, where the particle density vanishes{\footnote{Strictly speaking, there are two steady states,  \ie a no-particle state if the system starts with an even number of particles, and a one-particle state if it starts with an odd number of particles. However, these two steady states exhibit the same thermodynamic properties, so we identify them as the same state. Throughout this paper, states with identical thermodynamic properties will be identified.}}. It turns out that this dynamics has a universal late-time dynamics{\footnote{The universal late-time dynamics is sharply defined in the thermodynamic limit, just as most other universal quantities in a many-body system. In a finite system, such as a model under numerical simulations, this late-time dynamics appears in a time region where the density of the particles is much lower compared to the original density, while the total particle number is still much larger than 1. Below we will mostly speak of the universal late-time dynamics in the thermodynamic limit, and occasionally comment on its interpretation in a finite system.}}, which is characterized by, for example, the particle density as a function of time $\rho(t)$. More concretely, in one dimension (1d), $\rho(t)=\frac{1}{\sqrt{8\pi}}(Dt)^{-1/2}$ for large $t$, where $D$ is the diffusion constant. Here the prefactor $1/\sqrt{8\pi}$ is universal, \ie as long as this is an $A+A\rightarrow 0$ dynamics, this prefactor is independent of the microscopic details, such as the diffusion constant and the annihilation rate. The exponent $1/2$ in $(Dt)^{-1/2}$ appears {\it superuniversal}, in that it is not only independent of the microscopic details, but also applies to not only the $A+A\rightarrow 0$ dynamics, but also other types of dynamics, such as the $A+A\rightarrow A$ coagulation dynamics, or a hybrid of $A+A\rightarrow 0$ and $A+A\rightarrow A$ dynamics.

The setup of an $A+A\rightarrow A$ dynamics also consists of particles that can diffuse, just like the $A+A\rightarrow 0$ dynamics. However, it differs from the latter in that when two particles come together, at certain coagulation rate, they combine into a single particle of the same species that remains in the system, instead of becoming another species of particle that escapes from the system. In a hybrid of these two types of dynamics, when two particles come together, with a probability $p_{\AAtoO}$, they become another particle that escapes, and with a probability $p_{\AAtoA}=1-p_{\AAtoO}$, they combine into a single particle of the same type that stays in the system. Clearly, a pure $A+A\rightarrow 0$ dynamics and a pure $A+A\rightarrow A$ dynamics can be interpolated by a family of such hybrid dynamics, by tuning $p_{\AAtoA}$ from 0 to 1. For this entire family of dynamics, there is only a single steady state, \ie the state with a vanishing particle density. Interestingly, for each member of the family, there is a universal late-time dynamics, characterized by $\rho(t)\sim(Dt)^{-1/2}$ for large $t$ \cite{Henkel1995, Krebs1995, Simon1995, Henkel1996, Henkel2003}, where the prefactor varies continuously with the relative magnitude of the annihilation and coagulation rates of the limiting $A+A\rightarrow0$ and $A+A\rightarrow A$ dynamics (this relative magnitude can be tuned by tuning $p_{A+A\rightarrow 0}$), but is independent of the diffusion constant. We refer to this family of universal late-time dynamics a {\it universality family} of late-time dynamics, which is governed by the superuniversal $1/2$-exponent. To certain extent, this universality family possesses some resemblance to the notion of a conformal manifold in the context of conformal field theory, where a typical example is the infinitely many $(1+1)$-d compact-free-boson conformal field theories, which have identical central charge (counterpart of the $1/2$-exponent) but different compactification radii (counterpart of the prefactor) \cite{CFTBook}.

The reaction-diffusion dynamics studied in this paper is not of classical particles, but of Fibonacci anyons, a type of non-Abelian anyon that is not only interesting on its own, but also capable of performing universal quantum computation \cite{Nayak2007}. These Fibonacci anyons can emerge from certain $(2+1)$-d topological orders, and in such a realization our one dimensional system can be viewed as the interface between this topological order and the vacuum. In this setup, we ignore the exchange of anyons between bulk and the interface, and the reaction-diffusion dynamics we study here can be viewed as an effective model for the relaxation dynamics of these interface Fibonacci anyons from high temperature to low temperature, which consists of the diffusive motion of these anyons along the interface, and the anyon-reaction process induced by the couplings between these anyons and the environment. Alternatively, this dynamics may be realized by a hybrid quantum circuit exerted on a Fibonacci chain, where the diffusion and reaction can be realized by a stochastic unitary and nonunitary quantum channel acting on the anyons, respectively.

A significant difference between the reaction-diffusion dynamics of these anyons compared to that of classical particles is that these anyons are intrinsically quantum and nonlocal, in the sense that they should be described by a state in a Hilbert space, but this Hilbert space cannot be decomposed as a tensor product of local Hilbert spaces (see Appendix~\ref{app: review of Fibonacci} for a brief review of the basic structure of this Hilbert space). In fact, besides the position degree of freedom (DOF), these anyons carry a nonlocal internal DOF, due to which novel phenomena compared to the classical reaction-diffusion dynamics are expected. Clearly, in this dynamics there is still a single steady state with a vanishing density for the Fibonacci anyons. Our goal is to explore and characterize the universal late-time dynamics in this case.

Our work is partly motivated by Ref.~\cite{Nahum2019}, in which, among other subjects, the reaction-diffusion dynamics of Majorana defects in one dimension is studied. The Majorana defects therein can be viewed as emerging from the interface of segments of some $(1+1)$-d topological superconductors, and they also share some properties of non-Abelian anyons, such as the existence of the non-local internal DOF. And indeed, interesting phenomena of quantum origin were found there. For example, there is also a single steady state with a vanishing defect density, but its corresponding universal late-time dynamics has a density of the Majorana defects given by $\rho(t)=\frac{1}{\sqrt{2\pi}}(Dt)^{-1/2}$. Although the superuniversal $1/2$-exponent still shows up, the prefactor is twice of the classical $A+A\rightarrow 0$ dynamics. Furthermore, the internal DOF of the Majorana system can be characterized by its structure of quantum entanglement. We note that the underlying system of these Majorana defects, \ie the topological superconductors, can still be captured by a state in a Hilbert space that can be decomposed into tensor products of local (fermionic) Hilbert spaces, so the Fibonacci anyonic system studied here can be regarded to be more nonlocal than the Majorana defects. Accordingly, we will use a different approach to study the reaction-diffusion dynamics of Fibonacci anyons. In particular, we do not explicitly address the entanglement properties of the system, because the intrinsic non-locality of the Fibonacci anyons makes it tricky to define their entanglement in a physically motivated way (see, however, Refs. \cite{Bonesteel2006, Hikami2007, Fidkowski2008, Pfeifer2013, Kato2013, Bonderson2017} for recent development in this direction). More comparison between the Fibonacci and Majorana dynamics can be found in Sec. \ref{sec:discussions}.

Although there is a single steady state with a vanishing density of the Fibonacchi anyons, we find multiple universal late-time dynamics, corresponding to different initial conditions. We will use the anyon density and its two-point correlation function at late times to characterize the position DOF, and use the probability distribution of the fusion tree configuration (see below) to characterize the internal DOF. (In contrast, the internal DOF of the Majorana defects are characterized by their entanglement structure in Ref.~\cite{Nahum2019}.) Interestingly, at late times, we can identify regimes in which the internal DOF approaches one of two configurations to be described below, while the position DOF can be viewed as a hybrid of classical $A+A\rightarrow 0$ and $A+A\rightarrow A$ dynamics, with the probabilities of these two classical dynamics determined by the configuration of the internal DOF. So at late times the dynamics of both the internal and position DOF appear universal, and there is nontrivial interplay between classical and quantum behaviors. The nonlocal nature of the internal DOF makes this interplay especially intriguing.

The rest of the paper is organized as follows. 
In Sec.~\ref{sec:setup}, we describe the setups of the dynamics, where we detail the setup of pure-reaction dynamics in Sec.~\ref{subsec:pure-reaction} and the reaction-diffusion dynamics in Sec.~\ref{subsec:reaction-diffusion}.
In Sec.~\ref{sec:pure-reaction}, we study the pure-reaction dynamics without diffusion with  all $\tau$ initial fusion tree in Sec.~\ref{subsec:pure-reaction_alltau} and completely random initial fusion tree in Sec.~\ref{subsec:pure-reaction_random}. 
In Sec.~\ref{sec:reaction-diffusion}, we study the reaction-diffusion dynamics, where in Sec.~\ref{subsec:numerics} we present the numerical results and in Sec.~\ref{subsec:master_equation} we write down the master equation describing the dynamics and the mapping to the effective hybrid classical reaction-diffusion dynamics. Furthermore, In Sec.~\ref{subsec:2pt_corr}, we study the two-point correlation function, which provides further nontrivial justification of our mapping. And in Sec.~\ref{subsec:perturb_ini_diff}, we study the stability of the two initial fusion tree configurations. 
Finally, we discuss our results and outlooks in Sec~\ref{sec:discussions}.

\section{Setup}\label{sec:setup}

In this section, we describe the setup of the dynamics.
We will first consider the pure-reaction dynamics without diffusion.
A benefit of considering the pure-reaction dynamics is to gain some intuition for the dynamics of the internal fusion tree DOF by suppressing the diffusive motion of the anyons.
We will then consider the reaction-diffusion dynamics by allowing the anyons to perform random walks. Note that in all simulations in this paper, the initial particle number is definite.

\subsection{Pure-reaction dynamics}\label{subsec:pure-reaction}

\begin{figure}[h]
\centering
\includegraphics[width=0.48\textwidth]{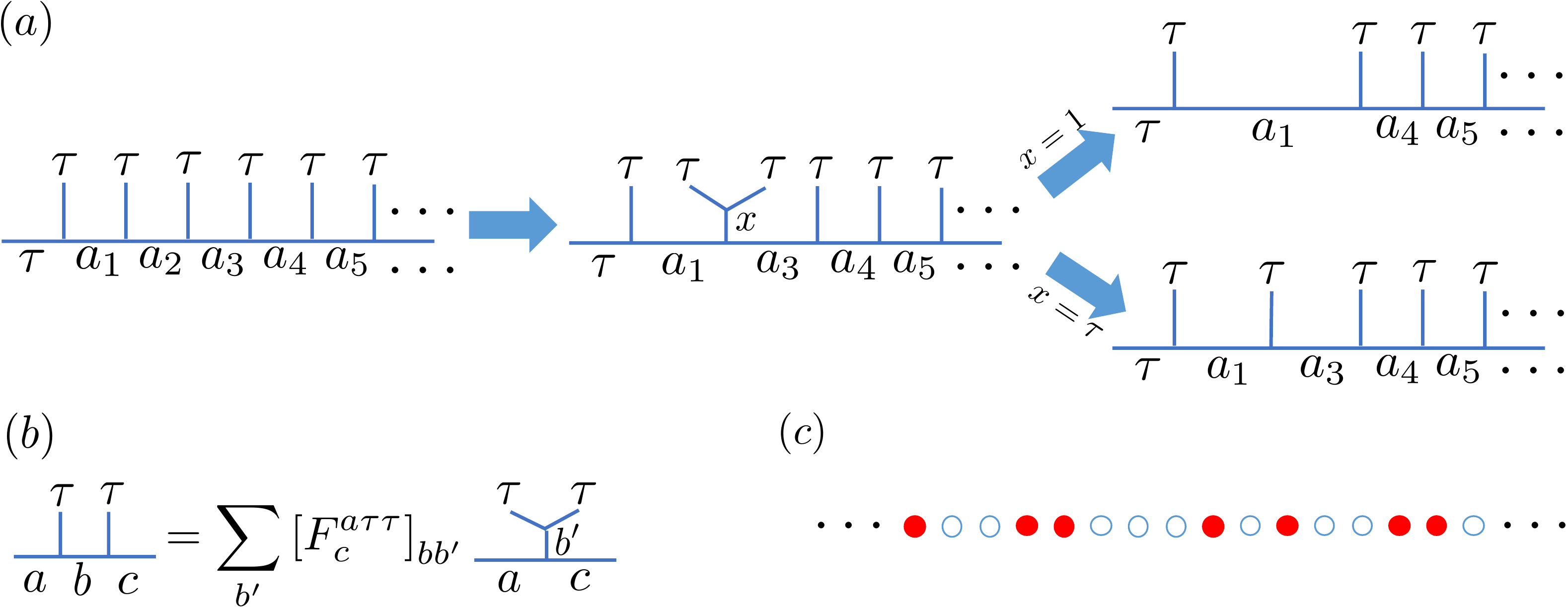}
\caption{Illustration of the setups. The left of (a) shows a state labeled by the fusion tree. Two of the Fibonacci anyons are measured, with outcome $x$. If $x=\trivial$, which necessarily implies $a_1=a_3$, this outcome is thrown away from the system. If $x=\tau$, this outcome is kept in the system. The probabilities to yield a particular outcome of a measurement is determined by the $F$-symbol as shown in (b).
(b) The fusion rule and the $F$-symbol.
(c) An example of a configuration of 16 contiguous sites of the lattice, where each red circle represents a site occupied by a Fibonacci anyon, and each empty circle represents an empty site.}
\label{fig: setup}
\end{figure}

In the setup of the pure-reaction dynamics, we consider an array of Fibonacci anyons with no diffusion.
The basis states of these Fibonacci anyons are labeled by a configuration of the fusion tree, \eg the states in Fig. \ref{fig: setup} (a) are labeled by the $a$'s in each segment of the fusion tree, and each of the $a$'s can either be $\trivial$, a trivial anyon, or $\tau$, a Fibonacci anyon. Due to the fusion rules, two contiguous segments cannot be simultaneously $\trivial$ (see Appendix \ref{app: review of Fibonacci} for a brief review of the basic physics of Fibonacci anyons). 
At each time step, two Fibonacci anyons corresponding to a pair of adjacent (vertical) fusion legs are chosen at random (with identical probabilities for all adjacent pairs), and their fusion product is measured.
The fusion product is given by the $F$-matrix as shown in Fig.~\ref{fig: setup}(b).
For example, as long as one of $a$ and $c$ is $\trivial$, then $b'$ is fixed. 
If $a=c=\tau$, $F^{\tau\tau\tau}_\tau=\left(\begin{array}{cc}\varphi^{-1} & \varphi^{-1/2}\\ \varphi^{-1/2} & -\varphi^{-1}\end{array}\right)$, where the first (second) row represents that $b=\trivial(\tau)$, and the first (second) column represents that $b'=\trivial(\tau)$. Here $\varphi=(\sqrt{5}+1)/2$ is the golden ratio.

\begin{figure}[h]
\centering
\includegraphics[width=0.49\textwidth]{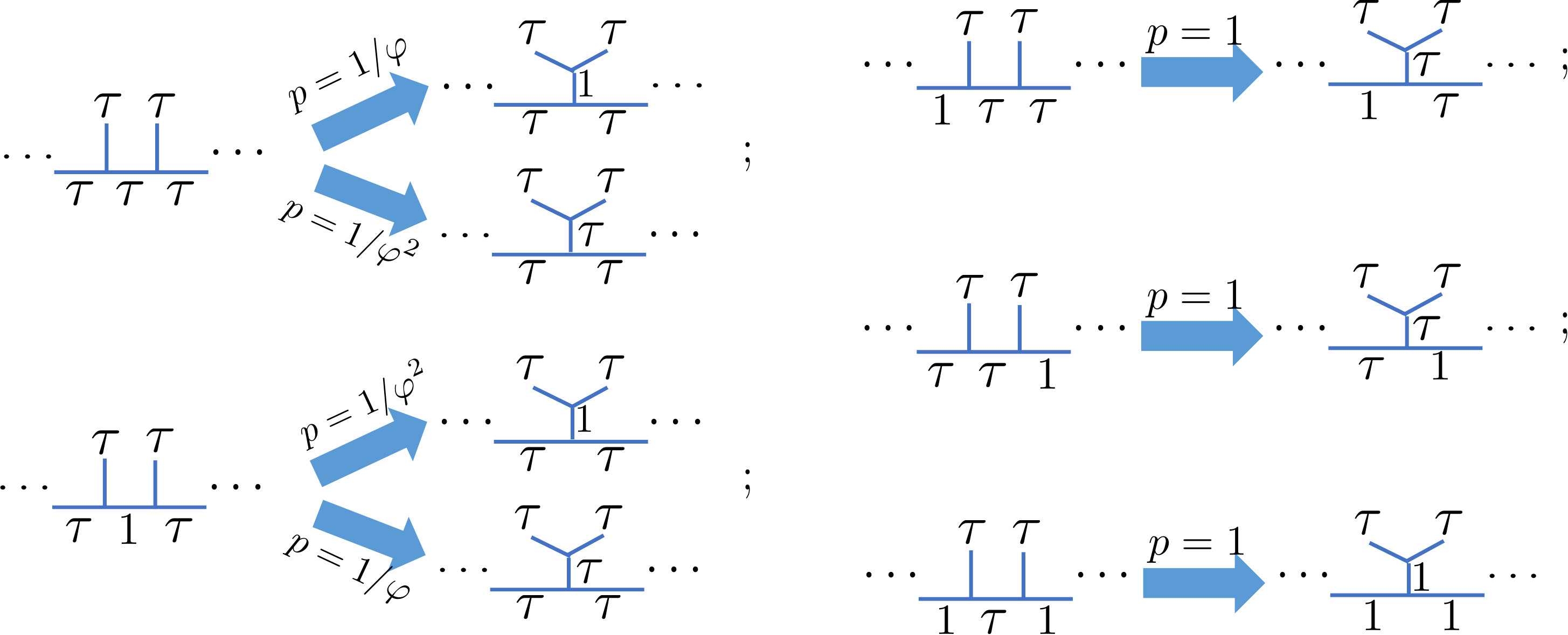}
\caption{The 5 types of length-3 segments, their measurement outcomes, and the corresponding probabilities.}
\label{fig: length-3 segments}
\end{figure}

Importantly, Fig.~\ref{fig: setup}(b) means that the left-hand side is a quantum superposition of the right-hand side. 
We then assume that the system is coupled to the environment in a way such that the pair of chosen adjacent anyons will be projected to a definite fusion measurement outcome of either $\trivial$ or $\tau$.
The associated probabilities are given in Fig.~\ref{fig: length-3 segments}, which are the squared values of the elements in the F-matrix, in accordance with Born's rule.
If the outcome is $\trivial$, then the measured anyons are annihilated, reducing the number of Fibonacci anyons by two; if the outcome is $\tau$, this anyon remains in the system and the number of Fibonacci anyons is decreased by one (see Fig.~\ref{fig: setup}). In a realization of the system in terms of a hybrid quantum circuit, this measurement-reaction protocol can in principle be designed by hand. In a realization at the interface between a topological order and the vacuum, this protocol is valid when the energy of a Fibonacci anyon is positive (in Sec.~\ref{sec:discussions} we will briefly discuss the case where a Fibonacci anyon has a negative energy). We stress that, in any case, this measurement-reaction process is local.

In our simulation steps, we keep track of the quantum trajectories without averaging over them. 
When evaluating a physical quantity, we first evaluate it for each quantum trajectory, and then average the results over the quantum trajectories.
While in principle, one should average over all the possible quantum trajectories, in practice, it is sufficient to sample over the quantum trajectories randomly with an enough number of realizations.
For physical quantities characterized by an operator that is not a function of the state of the system, the results obtained in this way are the same as obtained by first averaging the quantum trajectories to generate an ensemble of states, and then calculating the expectation value of this physical quantity with respect to this ensemble. 
This also means that no post-selection is needed to experimentally study these quantities.
A typical class of physical quantities not falling into this category are entanglement-related quantities, which have been widely studied recently in the context of hybrid quantum circuits \cite{Nahum2016, Nahum2017, Skinner2018, Li2018, Choi2019, Gullans2019}.

Within this setting, we can characterize the state of a system at a given time by $P(N, \{a_i\})$, the probability that the system has $N$ Fibonacci anyons with the segments of the fusion tree given by $\{a_i\}\equiv\{a_1, a_2, \cdots, a_N\}$. It will be useful to decompose this probability as $P(N, \{a_i\})=P_2(\{a_i\}|N)\cdot P_1(N)$, where $P_1(N)$ is the probability that the system has $N$ Fibonacci anyons, and $P_2(\{a_i\}|N)$ is the conditional probability distribution of the internal fusion tree DOF given a fixed $N$. We note that such a decomposition is unique for each $P(N, \{a_i\})$.

In this paper, one class of physical quantities we are interested in is the number or the density of anyons as time evolves. 
The above measurement-reaction protocol indicates that, to look at how fast the number of the Fibonacci anyons decreases, it is useful to consider the length-3 segments in the fusion tree associated with each pair of the adjacent Fibonacci anyons. 
For example, in the left-most figure of Fig.~\ref{fig: setup} (a), four length-3 segments are completely shown, corresponding to $\tau a_1a_2$, $a_1a_2a_3$, $a_2a_3a_4$, and $a_3a_4a_5$, respectively. All length-3 segments can be classified into 5 types, \ie $\tau\tau\tau$, $\tau\trivial\tau$, $\trivial\tau\tau$, $\tau\tau\trivial$ and $\trivial\tau\trivial$. 
In Fig.~\ref{fig: length-3 segments}, these length-3 segments are displayed together with their measurement outcomes and the corresponding probabilities.
Note that the instantaneous decay rate of the anyon number is given by
\beq \label{eq: decay rate master}
    r = p_1 + 2p_2~,
\eeq
where $p_1$ and $p_2$ are the probabilities of the measurement outcome being $\tau$ or $\trivial$, so the anyon number decreases by $1$ or $2$, respectively.
More explicitly,
\begin{align}\label{eqn:p1}
    p_1&= \varphi^{-2}p_{\tau\tau\tau}+\varphi^{-1}p_{\tau\trivial\tau}+p_{\tau\tau \trivial}+p_{\trivial\tau\tau} \\
    \label{eqn:p2}
    p_2&= \varphi^{-1}p_{\tau\tau\tau}+\varphi^{-2}p_{\tau\trivial\tau}+p_{\trivial\tau \trivial}~,
\end{align}
where $p_{\tau\tau\tau}$ is the probability of finding the $\tau\tau\tau$ segments, namely the number of $\tau\tau\tau$ segments in the fusion tree divided by the total number of length-3 segments. Other length-3 segment probabilities are defined similarly.

We therefore see that this instantaneous decay rate is controlled by the probability distribution of the length-3 segments, which is given by the probability distribution $P(N, \{a_i\})$ of the fusion tree at a given time. Take $p_{\tau\tau\tau}$ as an example. Suppose for a given fusion tree with $N\geqslant 3$ Fibonacci anyons and segment configuration $\{a_i\}$, the probability to find a $\tau\tau\tau$ segment is $p_{\tau\tau\tau}(N, \{a_i\})$. Then $p_{\tau\tau\tau}=\sum_{N\geqslant 3, \{a_i\}}p_{\tau\tau\tau}(N, \{a_i\})\cdot P(N, \{a_i\})$. Note that $p_{\tau\tau\tau}(N, \{a_i\})$ is purely determined by the specific fusion tree and is independent of the probability distribution of different fusion trees, characterized by $P(N, \{a_i\})$. More generally, for a length-$\ell$ segment labeled by $\{a_i\}=\{a_1, a_2, \cdots, a_\ell\}$, its probability in a state characterized by a probability distribution $P(N, \{a'_j\})$ is
\begin{equation}
p_{\{a_i\}}=\sum_{N\geqslant\ell, \{a'_j\}}p_{\{a_i\}}(N, \{a'_j\})P(N, \{a'_j\})~,
\end{equation}
where $p_{\{a_i\}}(N, \{a'_j\})$ is the probability to find a length-$\ell$ segment $\{a_i\}$ in a length-$N$ fusion tree with segments $\{a'_j\}$. It can be shown that knowing the ensemble-averaged probability $p_{\{a_i\}}$ for all length-$\ell$ segments is sufficient to uniquely determine the ensemble-averaged probabilities for all segments with shorter lengths.

\subsection{Reaction-diffusion dynamics}\label{subsec:reaction-diffusion}

After gaining some intuition for the dynamics of the internal fusion tree DOF from the pure-reaction dynamics, we will move to the main subject: the reaction-diffusion dynamics. We consider a one dimensional lattice with $L$ sites, where each site is either empty or occupied by a single Fibonacci anyon (see Fig.~\ref{fig: setup}~(c)). 
Ignoring the empty sites, these Fibonacci anyons just form the array discussed in the pure-reaction process, and their basis states are again labeled by the fusion tree as before. 
If two Fibonacci anyons are at two adjacent sites, with probability $p_R$ ($0\leqslant p_R\leqslant 1$), they undergo the measurement-reaction process as described in the pure-reaction dynamics, where if the fusion outcome of two Fibonacci anyons is $\tau$, this $\tau$ can occupy the site of either of the two original anyons, with equal probability. Here $p_R$ is introduced to control the reaction rate. 
After the measurement-reaction protocol, the Fibonacci anyons diffuse, \ie each of them moves with a probability $\frac{1}{2}p_D$ to either the left or right by one lattice spacing (if that neighboring site is empty), and stays still with a probability $1-p_D$, where $0\leqslant p_D\leqslant 1$ is a proxy of the diffusion constant. The combination of the above reaction and diffusion processes counted as one time step in our simulation.

If more than two anyons occupying some contiguous sites, we group the pairs starting from the left to perform the measurements. 
For example, if $5$ anyons happen to be occupying the contiguous sites, then we group pairs ($1$,$2$) and ($3$,$4$) for the measurements.
We note that the probability of these events are extremely low if the anyon density is low, and we expect that the detailed implementation of such short-distance processes does not affect the universal late-time dynamics.
Notice that we will ignore the braiding processes of the Fibonacci anyons. In the realization on the boundary of a $(2+1)$-d topological order, this is valid if these anyons have strong enough interactions to ensure that the reaction occurs before the braiding; and in the realization of a hybrid quantum circuit, braiding can be simply forbidden by hand. Even if braiding is included, in this 1d system, it is not expected to alter the universal late-time dynamics of our interest because i) at late times the typical distance between anyons is large, so the most important processes involving braiding occur only between a pair of anyons, without touching a third anyon; and ii) braiding two anyons will only induce an unimportant phase factor depending on their fusion outcomes.

Again, one physical quantity we are interested in is the number, or equivalently, the density $\rho(t) \equiv N(t)/L$ of the anyons as a function of time averaged over the realizations of the quantum trajectories.
In fact, it is a very general feature for any 1d reaction-diffusion process to have $\rho(t) \sim (Dt)^{-1/2}$ behavior at late times, where $D$ is the diffusion constant \cite{Bramson1980, Torney1983, Lushnikov1987, Toussaint1983, Henkel1995, Henkel1996, Henkel2003, Krebs1995, Simon1995}.
A heuristic argument for this behavior can be found in various references \cite{Toussaint1983, Tauber2005, Nahum2019}. Here we repeat the argument for readers' convenience. 
In the dilute limit $\rho \ll 1$, the typical particle spacing is $\ell \sim 1/\rho$. The time scale for particles to diffuse to each other is $\Delta t \sim \ell^2/D \sim 1/(D\rho^2) $. On the other hand, the change of the particle density is proportional to the density itself $\Delta\rho \sim -\rho$. These two conditions give us $\Delta \rho/\Delta t \sim -D\rho^3$, or $\rho(t)\sim (Dt)^{-1/2}$, which explains the 1/2-exponent. Remarkably, not only the exponent $1/2$, but also the dimensionless coefficient in front of this $(Dt)^{-1/2}$ shows universality, \ie they are to certain extent independent of the microscopic details.
It is, however, not always easy to obtain this coefficient exactly.
As we will see, for some special initial fusion trees, the dynamics of the position DOF can be mapped to a hybrid of classical $A + A \rightarrow 0$ and $A + A \rightarrow A$ processes at late times, where the coefficient of the $\rho(t) \sim (Dt)^{-1/2}$ at long time can be obtained exactly.

\section{Pure-reaction dynamics}\label{sec:pure-reaction}

In this section, we study and present the results of the pure-reaction dynamics. Generically, one expects that the late-time dynamics depends on the initial state of the internal DOF, potentially in some complex fashion. So it is useful and interesting to identify some special initial condition of the internal DOF, in which the dynamics takes a simpler form.
We will consider two specific types of the initial fusion tree: the ``all-$\tau$" and ``completely random" configurations, both with a definite initial number of Fibonacci anyons.
The all-$\tau$ is a configuration where all segments of the fusion tree are $\tau$, while the completely random is a configuration drawn from all possible fusion trees completely randomly with a uniform probability. 
As we will see, these two fusion trees correspond to ``internally steady states" for the pure-reaction process (which roughly means that the probability distribution of the internal DOF, $P_2(\{a_i\}|N)$, is time independent, but see below for the more refined definitions of internally-steady states at level-$\ell$), and they are unstable and stable solutions, respectively. 

In our numerical simulations, we generate an initial fusion tree and run the measurement protocol for $1000$ measurement realizations, starting with $N_0=2 \times 10^5$ anyons. 
At each time step, we randomly choose two adjacent anyons in the fusion legs and implement the measurement-reaction process. We then monitor the number of anyons as a function of time $N(t)$.

\subsection{All-$\tau$ initial fusion tree}\label{subsec:pure-reaction_alltau}

The all-$\tau$ state is perhaps the simplest fusion tree, where each segment of the fusion tree is $\tau$. With this initial condition, although a nontrivial (binomial) distribution of $P_1(N)$ will be generated as time evolves, it is clear that each segment of the fusion tree will always be $\tau$. So we identify the all-$\tau$ state as an {\it internally super-steady state}, in the sense that $P_2(\{a_i\}|N)$ is time independent, for {\it all} $N$. This is to be contrasted with an {\it internally level-$\ell$-steady state}, for which the probability to find a given length-$\ell$ segment among all length-$\ell$ segments is time independent. For example, if $p_{\tau\tau\tau}, p_{\tau\trivial\tau}, p_{\tau\tau\trivial}, p_{\trivial\tau\tau}, p_{\trivial\tau\trivial}$ are time independent, then the state is internally level-3-steady. Note that an internally level-$\ell_1$-steady state is necessarily internally level-$\ell_2$ steady, for all $\ell_2<\ell_1$. Also, in the thermodynamic limit an internally super-steady state can be viewed as an internally level-$\infty$-steady state.\footnote{In a finite system, such as the ones being simulated here, suppose at time $t$ the maximal possible anyon number is $N_{\rm max}(t)$, where $N_{\rm max}(t)$ is finite and decreases as time evolves, then being internally super-steady (over a period $T$) should be interpreted as being internally level-$N_{\rm max}(T)$-steady for all $t\leqslant T$.}

\begin{figure}
    \includegraphics[width=\columnwidth]{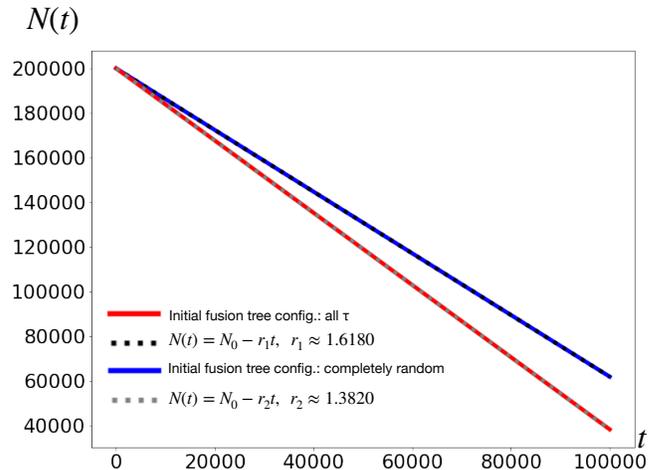}
    \caption{The number of Fibonacci anyons as a function of time in the pure-reaction dynamics. 
    The initial fusion tree configurations are prepared with the all -$\tau$ and completely random configuration.
    The analytical predictions of the decay rates well describe the numerical results. 
    The numerical calculations are averaged over $1000$ measurement realizations.
    }
    \label{fig:N_t}
\end{figure}

\begin{figure}[h!]
    \includegraphics[width=\columnwidth]{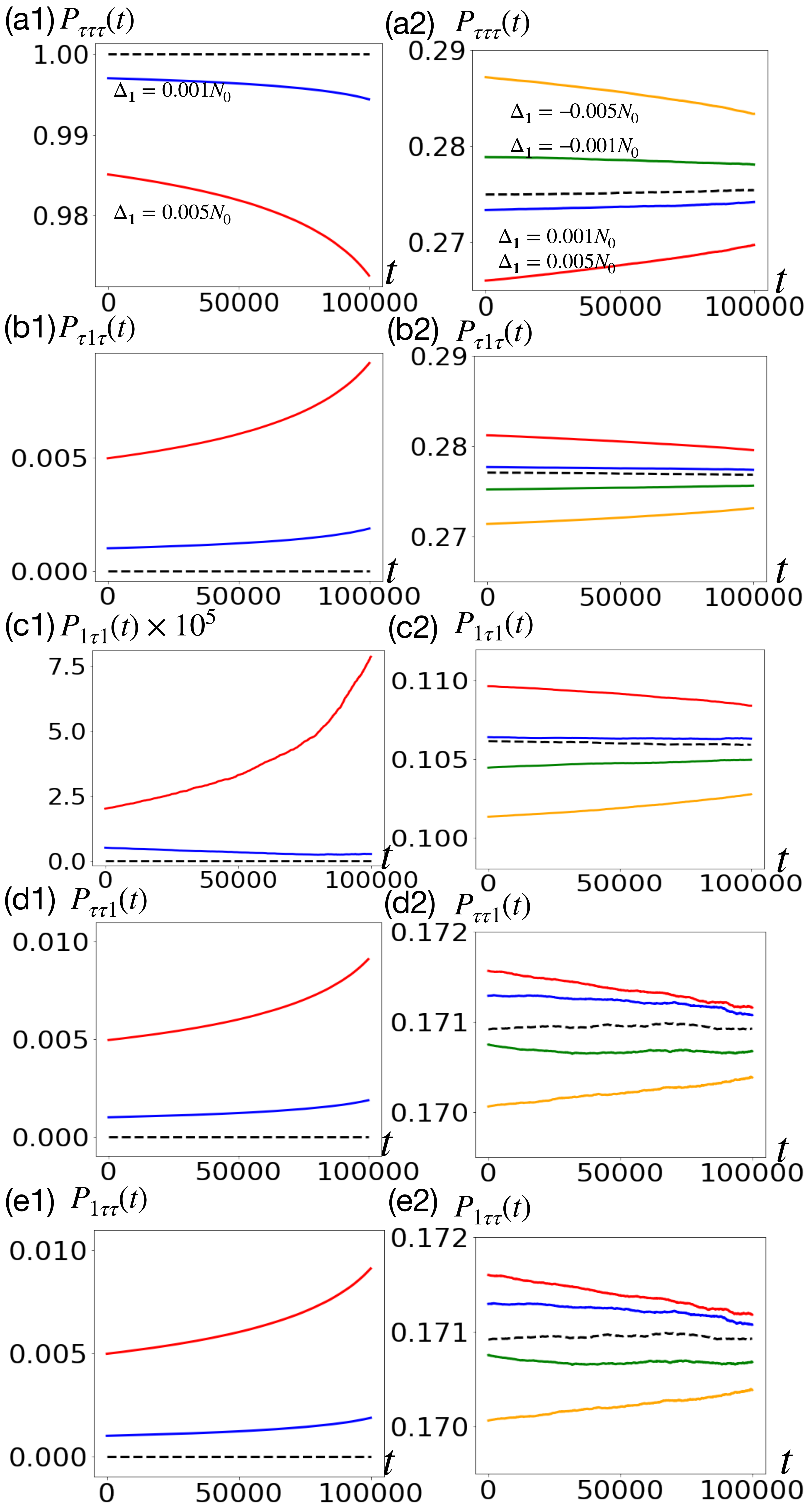}
    \caption{Left panel (a1)-(e1): The distributions of the length-3 segments in the pure-reaction dynamics with the all-$\tau$ initial fusion tree and its perturbation $\Delta_\trivial$, the deviation of the number of $\trivial$'s in the initial fusion tree with $N_0 =2 \times 10^5$ initial anyons. The results of the perturbation suggests that all-$\tau$ configuration is unstable. 
    Right panel (a2)-(e2): Similar to the left panel but with the completely random initial fusion tree and its perturbations $\Delta_\trivial$.
    The results of the perturbed initial fusion tree suggest that the completely random configuration is likely stable.
    All calculations are averaged over $1000$ measurement realizations.
    }
    \label{fig:Fibbpert}
\end{figure}

In Fig.~\ref{fig:N_t}, we can see that the number of anyons $N(t)$ decays linearly as a function of time, with a rate
\beq \label{eq: decay rate all-tau no diffusion}
r_1=\varphi^{-2}+2\varphi^{-1}\approx 1.618~.
\eeq
We remark that a time-independent decay rate is a property of an internally level-3-steady state. Notice $r_1$ is the rate {\em per time step}, and it is appropriate to view it as being dimensionless. 
This rate can be obtained as the following.
When we perform the measurement at each time step, the underlying fusion tree for any adjacent anyons is always $\tau\tau\tau$. 
According to our measurement-reaction protocol, the probabilities of annihilating one anyon and two anyons are $p_1=\varphi^{-2}$ and $p_2=\varphi^{-1}$, respectively.
Using Eq.~\eqref{eq: decay rate master}, we therefore obtain the decay rate in Eq.~(\ref{eq: decay rate all-tau no diffusion}), at all times.
Indeed, this rate agrees with the numerical simulation as shown in Fig.~\ref{fig:N_t}.
The internally-steady state property of the all-$\tau$ fusion tree can be further verified numerically in Fig.~\ref{fig:Fibbpert}(a1)-(e1), where the probabilities of all the length-3 segments are constant in time, denoted by the black dashed lines. Note that these probabilities are averaged over $1000$ measurement realizations at a given time.

It is interesting to ask about the stability of this internally-steady state, \ie if the initial condition is changed slightly so that certain segments of the fusion tree are $\trivial$, under the dynamics whether the system evolves toward a state where all segments are $\tau$. If so (not), this internally-steady state is stable (unstable).

To this end, we consider a perturbation of the initial fusion tree where we mutate a small number of $\tau$'s in the all-$\tau$ initial fusion tree to $\trivial$. We randomly choose $\Delta_\trivial$ number of $\tau$'s in a way where the fusion rules are still satisfied.
In this case, at least initially, for most of the time the length-3 segment with $\tau\tau\tau$ will be measured, which decreases the number of $\tau$-segments. With a small probability, other length-3 segments may also be measured, which causes a decrease of either the $\trivial$-segments or $\tau$-segments. Whether the all-$\tau$ state is stable is then determined by the competition of two factors: (i) the large initial number of the $\tau$-segments, and (ii) the large initial decay rate of the $\tau$-segments.

The result of this competition in the dilute limit of $\trivial$'s can be understood as follows. Denote the number of the $\tau$-segments and $\trivial$-segments at time $t$ by $N_\tau(t)$ and $N_\trivial(t)$, respectively. Note that these are the numbers of the segment in the fusion tree, not to be confused with the number of physical anyons $N(t)=N_{\tau}(t)+N_\trivial(t)$.
We will see that $N_\trivial(t)/N_\tau(t)$ increases as $t$ increases, which shows that the all-$\tau$ state is unstable.

To this end, we can write down the following approximate continuous-time rate equations for $N_\tau(t)$ and $N_\trivial(t)$ that are valid in the dilute limit,
\beq
\begin{split}
&\frac{dN_\tau}{dt}=-\tilde r_0\\
&\frac{dN_\trivial}{dt}=-\frac{N_\trivial}{N_\tau}\frac{1}{t_0}~,
\end{split}
\eeq
where $t_0$ can be viewed as the duration of each time step of the numerical simulation, and $\tilde r_0\equiv r_0/t_0$ is the rate with a dimension of the inverse time. In the first equation above, the decrease of the number of the $\tau$-segments is completely attributed to the length-3 segments with $\tau\tau\tau$, and the probability to find such a length-3 segment is approximately unity because $N_{\tau}(t)/N(t) \approx 1$. In the second equation above, the probability of finding a length-3 segment with $\tau \trivial\tau$, $N_\trivial/N$, is approximated by $N_\trivial/N_\tau$. These approximations are valid if the all-$\tau$ state is stable.

These rate equations can be solved, yielding
\beq
\begin{split}
    &N_\tau(t)=N_\tau(0)-\tilde r_0t\\
    &N_\trivial(t)=N_\trivial(0)\left(\frac{N_\tau(0)-\tilde r_0t}{N_\tau(0)}\right)^{\frac{1}{r_0}}~,
\end{split}
\eeq
which implies the ratio of the numbers of the $\trivial$-segments and $\tau$-segments is
\beq \label{eq: ratio of numbers}
\frac{N_\trivial(t)}{N_\tau(t)}=\frac{N_\trivial(0)}{N_\tau(0)^{\frac{1}{r_0}}}\cdot\frac{1}{\left(N_\tau(0)-\tilde r_0t\right)^{1-\frac{1}{r_0}}}~.
\eeq
This ratio increases with time, which means that the all-$\tau$ state is unstable. We have checked that Eq. \eqref{eq: ratio of numbers} agrees with the numerical results well (not shown).
This instability of the solution is also supported numerically in Fig.~\ref{fig:Fibbpert}(a1)-(e1), where the probabilities of the length-3 segments flow away from the all-$\tau$ distribution upon perturbations of $\Delta_\trivial$.

\subsection{Completely random initial fusion tree}\label{subsec:pure-reaction_random}

Knowing that the all-$\tau$ state is internally steady but unstable, it is interesting to ask if this system has any stable internally steady state at all, and which initial conditions will approach this internally steady state as time evolves.

We find that there is another internally steady state, the completely random state, and there is evidence that this steady state is stable under small perturbations. In an array with $N$ Fibonacci anyons, there are $F_N$ allowed states (see Appendix \ref{app: review of Fibonacci} for the explicit expression of $F_N$), and each of them has an equal probability in the completely random state. This completely random state can be physically realized by tuning the effective temperature of the Fibonacci anyons to infinite.

We simulate the system as follows.
We prepare an initial fusion tree configuration drawn from the completely random ensemble 
(see Appendix \ref{app: completely random state} for an efficient procedure to generate the completely random initial fusion tree ensemble).
At each time step, we randomly choose a nearest-neighbor pair of anyons and perform the measurement-reaction procedure, determining if it fuses to $\tau$ or $\trivial$.
For the same initial fusion tree, we repeat the calculation with $1000$ measurement realizations.  
We monitor the average number of anyons as a function of time, $N(t)$, as shown in Fig.~\ref{fig:N_t}.
Note that here we only use one initial fusion tree configuration drawn from the completely random ensemble. 
However, for the quantities we are interested in, we observe that such a configuration indeed has the property of self-averaging. That is, it reproduces the result from the completely random ensemble average very well.
While not shown in the paper, we have tested different initial fusion trees drawn from the completely random ensemble, and the results all agree with each other within statistical errors. 

We can see from Fig.~\ref{fig:N_t} that the anyon number $N(t)$ decays linearly, suggesting that the completely random distribution is internally steady with at least level-3. This can indeed also be seen and supported by the length-3 distributions as a function of time shown in Fig.~\ref{fig:Fibbpert}(a2)-(e2) denoted by the black dashed lines, where the small fluctuations in time are likely due to statistical error.
The decay rate is again given by Eqs.~(\ref{eq: decay rate master}),(\ref{eqn:p1}) and (\ref{eqn:p2}), and the probabilities of the five length-3 segments for the completely random state are computed in Appendix~\ref{app: completely random state}.
This gives us $p_1=(3\varphi+1)/(4\varphi +3)$ and $p_2 =(\varphi+2)/(4\varphi+3)$. 
Therefore, the decay rate 
\beq \label{eq: r2}
r_2=(5\varphi+5)/(4\varphi+3) \approx 1.3820~,
\eeq
which agrees with our numerical result.

We conjecture that the completely random state is actually internally super-steady, which can be justified (but not rigorously proved) as follows. Denote the total number of Fibonacci anyons in the array by $N(t)$. Using Eq. \eqref{eq: decay rate master}, we get an rate equation of the total number of Fibonacci anyons $N$,
\begin{align}
\frac{\Delta N}{\Delta t}=&-\left(2\varphi^{-1}+\varphi^{-2}\right)p_{\tau\tau\tau}-\left(2\varphi^{-2}+\varphi^{-1}\right)p_{\tau 1\tau}\notag \\
&-p_{1\tau\tau}-p_{\tau\tau 1}-2p_{1\tau 1}~.
\end{align}
Analogously, we get the rate equations for the numbers of the length-3 segments,

\beq \label{eq: rate equations length-3}
\begin{split}\notag
    \frac{\Delta N_{\tau\tau\tau}}{\Delta t}&=-p_{\tau\tau\tau}(1+\varphi^{-1})+p_{1\tau\tau\tau 1}\varphi^{-1}\\
    &-2p_{\tau 1\tau\tau\tau}+p_{\tau\tau 1\tau\tau}+p_{\tau 1\tau}\varphi^{-1}-p_{1\tau 1\tau 1}\varphi^{-1}\\
    \frac{\Delta N_{1\tau\tau}}{\Delta t}&=-p_{1\tau\tau\tau 1}\varphi^{-1}-2p_{\tau 1\tau\tau 1}-p_{\tau\tau 1\tau\tau}+p_{1\tau 1\tau 1}\varphi^{-1}\\
    \frac{\Delta N_{1\tau 1}}{\Delta t}&=p_{1\tau\tau\tau 1}\varphi^{-1}+2p_{\tau 1\tau\tau 1}-(p_{1\tau 1}+p_{\tau 1\tau})\\
    &+p_{\tau\tau 1\tau\tau}-p_{1\tau 1\tau 1}\varphi^{-1}\\
    \frac{\Delta N_{\tau 1\tau}}{\Delta t}&=-(p_{1\tau 1}+p_{\tau 1\tau})~,
\end{split}
\eeq
where $p_{1\tau\tau\tau 1}$ is the probability of finding a length-5 segment with $1\tau\tau\tau 1$ and similar symbols have similar meanings. In the above, we have also used the fact that $p_{1\tau\tau}=p_{\tau\tau 1}$ for $N\gg 1$. From these rate equations, we see that the decay rate of a type of segment with certain length depends on the probabilities of some segments with longer lengths.

In an internally-steady state, the probabilities such as $p_{\tau\tau\tau}$ should be time independent, which requires that $\frac{\Delta N_{\tau\tau\tau}/\Delta t}{\Delta N/\Delta t}=p_{\tau\tau\tau}$, etc. Substituting the results of the probabilities $p_{\tau\tau\tau}$ etc from Appendix \ref{app: completely random state} into the above rate equations, it is straightforward to check that relations like $\frac{\Delta N_{\tau\tau\tau}/\Delta t}{\Delta N/\Delta t}=p_{\tau\tau\tau}$ indeed hold. 
This observation supports that the completely random state is an internally level-3-steady state.
While in principle, one can write down the rate equations for the longer-length segments and verify the time independence of them, the task becomes formidable. We therefore only consider the rate equations up to length-3 segments. However, as we observed above, the rate equation for the number of a certain segment involves the number of some longer-length segments. In order for the completely random state to be internally level-3-steady, it is natural that it is actually internally level-5-steady, since the probabilities of some length-5 segments enter Eq. \eqref{eq: rate equations length-3}. Reasoning in a similar fashion, it appears natural that the completely random state is in fact internally super-steady in the thermodynamic limit. That is, in the thermodynamic limit, at all times for all $N$, $P_2(\{a_i\}|N)$ is given by the completely random probability distribution, as long as the initial fusion tree distribution is completely random.\footnote{In a finite but large system, we expect that over a period $T$ the completely random state is (at least approximately) internally level-$\ell$-steady, where $\ell$ satisfies $1\ll\ell\lesssim N_{\rm max}(T)$. When the number of remaining anyons is small, this internally-steady-state nature and the linear decrease of the anyon number can be violated.}

In passing, we note that the above argument actually suggests that all internally steady states in this dynamics are likely to be internally super-steady (in the thermodynamic limit), although at this stage it is unclear whether there are other internally steady states.

Next, we turn to the stability of the completely random state. Analyzing the stability from the rate equations is in fact challenging for the completely random state, unlike the case of the all-$\tau$ state.
Here, we test the stability numerically.
We perturb the initial fusion tree configuration by randomly changing $\Delta_\trivial$ number of $\tau$'s into $\trivial$'s if $\Delta_\trivial > 0$ or $|\Delta_\trivial|$ number of $\trivial$'s into $\tau$'s if $\Delta_\trivial < 0$, in a way that the fusion tree is physically allowed.
In Fig.~\ref{fig:Fibbpert}(a2)-(e2), we show the probability of the length-3 segments as a function of time. 
It appears that the probability distributions flow to the completely random distribution, and we therefore conclude that such a distribution is likely a stable internally super-steady state.

It seems the perturbed distributions never reach any final steady states within our simulation. It is therefore natural to ask if one can see them to do so. 
While we have tried the simulations with a larger number of initial particles $N_0$ so that we can extend our simulation time steps, we find that the transient time scale is likely to be a function of $N_0$, and is larger than the decay time scale of the the particle number. Therefore, it is likely impossible to see the length-3 distributions to reach their final steady states with simulations: the anyon number will reach zero before the length-3 distributions reach their steady state (if any).

To summarize this section, we have studied the pure-reaction dynamics of the Fibonacci anyons, with the initial state being an all-$\tau$ or completely random state with a fixed initial anyon number. We have shown that the all-$\tau$ state is an unstable internally super-steady state, and we have argued that the completely random state is a stable internally super-steady state. This suggests that the reaction processes tend to drive the internal DOF of the system to the completely random state, which further motivates us to use this state as an initial condition for the internal DOF in the reaction-diffusion dynamics in the next section.

\begin{figure}[h]
    \includegraphics[width=\columnwidth]{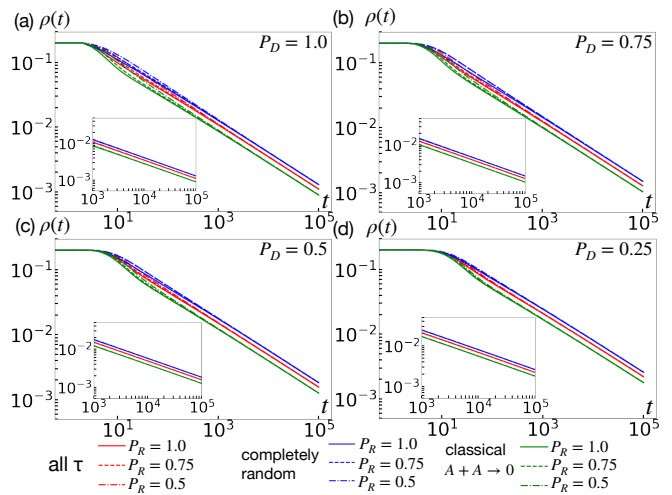}
    \caption{(a)-(d)The density $\rho(t)$ of the reaction-diffusion process of the Fibonacci anyons and the classical $A+A \rightarrow$ process with varying diffusion probability $P_D$ and reaction probability $P_R$.
    For all the processes, the long time behavior of the density show $\rho(t) \sim t^{-1/2}$ as expected. 
    The collapse of the curves for different $P_R$'s shows the irrelevance of the reaction probability in the late-time dynamics.
    Insets: Zoomed-in scale for the data. The parallels further verify the expected Fibonacci result to be $\rho(t)=c \rho_{A+A\rightarrow 0}(t)$ with some constant $c$. 
    }
    \label{fig:diffusion1}
\end{figure}

\begin{figure}
    \includegraphics[width=\columnwidth]{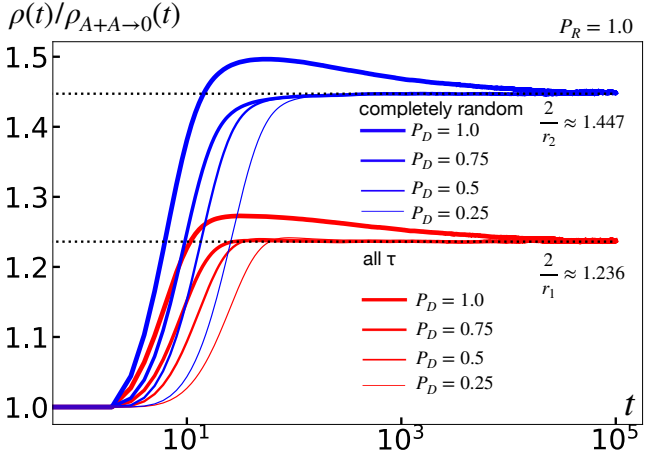}
    \caption{The ratio of the density of Fibonacci anyons in the reaction-diffusion dynamics to the result of the classical $A+A \rightarrow 0$ dynamics,
    with initial Fibonacci fusion tree as all-$\tau$ or completely random, with varying diffusion probability $P_D$.
    At late times, the ratio approaches $2/r_1$ and $2/r_2$ for the all-$\tau$ and completely random initial fusion tree, respectively. These ratios are the same for different $P_D$.
    }
    \label{fig:diffusion2}
\end{figure}

\section{Reaction-diffusion processes}\label{sec:reaction-diffusion}

\subsection{Numerical results}\label{subsec:numerics}

Now we move to the reaction-diffusion dynamics.
Consider a chain with $L$ sites having either $\tau$ or $\trivial$ (no anyon) and an underlying fusion tree configuration associated with it.
At each time step, each anyon has a probability $P_D$ to perform an unbiased random walk. Namely, it has probability $P_D/2$ to hop to the left or right if the site to be hopped on is empty.
After the random walk, if there are pairs of the anyons happening to be adjacent, there is a probability $P_R$ that a reaction occurs on these pairs of the anyons as described in Sec~\ref{sec:setup}.
Here $P_D$ and $P_R$ are introduced to tune the diffusion and reaction rates, respectively. Motivated by the discussion in Sec~\ref{sec:pure-reaction},
in our numerical simulation, we will first prepare a definite number $N_0$ of anyons to be equally spaced initially and with the initial fusion tree configuration prepared as all-$\tau$ or drawing from the completely random ensemble. Later we will also perturb the initial conditions. Again we are interested in the density of anyons as a function of time and the probability distribution of the internal fusion tree DOF.

As discussed in Sec.~\ref{sec:setup}, we expect $\rho(t) =c \frac{1}{\sqrt{8 \pi D t}}$ at long times.
To extract the prefactor $c$, it is more convenient and customary to compare the result to the classical $A\!+\!A\! \rightarrow\! 0$ dynamics, where $\rho_{A\!+\!A\! \rightarrow\! 0}(t) = \frac{1}{\sqrt{8\pi D t}}$ can be exactly solved \cite{Bramson1980, Torney1983, Lushnikov1987}.
We therefore expect $\rho(t)=c\rho_{A\!+\!A\! \rightarrow\! 0}(t)$ and can extract the constant $c$ from the ratio of the numerical results of $\rho(t)$ and $\rho_{A+A\rightarrow 0}(t)$, without explicitly extracting the diffusion constant $D$. 

In Fig.~\ref{fig:diffusion1}, we show the results of the simulations for the Fibonacci dynamics for different parameters $P_D$ and $P_R$, together with the classical $A\!+\!A\! \rightarrow\! 0$ results for comparison.
All the simulations are performed with $L=10^6$, $N_0=2 \times 10^5$ and averaged over $1000$ realizations of the reaction-diffusion dynamics for a given initial condition.
While it is well-established that $P_R$ does not affect the long-time dynamics in the classical $A\!+\!A\! \rightarrow\! 0$ dynamics, the collapse of the curves in Fig.~\ref{fig:diffusion1}(a)-(d) also confirms this aspect in the Fibonacci dynamics.
The figures also verify the expected behavior $\rho(t)=c\frac{1}{\sqrt{8\pi D t}}$.

To determine its prefactor $c$, we take the ratio of the particle density in the Fibonacci dynamics to that in the classical $A+A \rightarrow 0$ dynamics, as shown in Fig.~\ref{fig:diffusion2}.
We can see that the ratio $c$ is independent of $P_D$ from the collapse of the curves at long times. These findings imply that the late-time dynamics associated with an initial all-$\tau$ and completely random fusion tree distributions are universal, and they are also distinct, although they correspond to a single steady state.
We find the ratios to be close to $2/r_1$ and $2/r_2$ for the all-$\tau$ and completely random initial fusion tree configuration, with $r_1$ and $r_2$ given by Eqs. \eqref{eq: decay rate all-tau no diffusion} and \eqref{eq: r2}, respectively. These ratios can in fact be obtained exactly, as we illustrate in the next subsection.

\subsection{Master equation}\label{subsec:master_equation}

To understand the late-time dynamics of $\rho(t)$, especially the ratio of it to the classical $A+A \rightarrow 0$ dynamics, we use a continuous-time master equation to describe the dynamics. In general, denote the (time-dependent) probability for a system to be in a state $s$ by $P_s$, the master equation takes the form
\beq
\frac{\partial P_s}{\partial t}=\sum_{s'\neq s}(T_{s'\rightarrow s}P_{s'}-T_{s\rightarrow s'}P_s)~,
\eeq
where $T_{s_1\rightarrow s_2}$ represents the rate for the system to go from state $s_1$ to state $s_2$. The physical meaning of this master equation is clear: the change of the probability in a state is the difference between the rate to go from other states to this state and the rate to go from this state to other states.

To apply the master equation approach to our setup, at each site $i$, we assign the occupation number of the anyon $n_i =0$ or $1$ representing empty or occupied site respectively, and the fusion tree segment $a_i = \trivial $ or $\tau$, denoting the fusion label between site $i$ and $i+1$. Since we track the anyon occupation number, the fusion tree $\{a_i \}$ here is the augmented ``second-quantized" version of the fusion tree discussed in the previous sections. 
In particular, in addition to requiring that $a_i$ and $a_{i+1}$ cannot simultaneously be  $\trivial$ when $n_{i+1}=1$, we will also need $a_{i+1}=a_i$ when $n_{i+1}=0$ to have a consistent fusion tree. Here, while we formally allow any configurations $\{n_i, a_i\}$, the unphysical configurations (namely configurations not consistent with the fusion rule) will have zero probabilities.
Denoting the probability of a system in a state $\{n_i, a_i\}$ by $P(\{ n_i,a_i\};t)$, we can write down a master equation that models the Fibonacci reaction-diffusion dynamics as 
\beq \label{eq: master equation}
\frac{\partial P(\{ n_i, a_i\};t)}{\partial t}=L_D(P)+L_R(P)~,
\eeq
where $L_D(P)$ and $L_R(P)$ represent the contribution from diffusion and reaction, respectively. 
They are given by
\begin{widetext}

\beq
\begin{split}
L_D(P)=D\sum_i\Big[&P\left(\cdots,\! \begin{array}{c}n_{i-1},\\ a_{i},\end{array}\! \begin{array}{c}1,\\ a_{i+1}, \end{array}\! \begin{array}{c}0, \\ a_{i+1},\end{array}\! \cdots; t\right)\delta_{n_i,0}\delta_{n_{i+1},1}\delta_{a_{i-1},a_i}-P(\{n_i, a_{i}\}; t)\delta_{n_i,0}\delta_{n_{i+1},1}\delta_{a_{i-1},a_i}\\
+&P\left(\cdots,\! \begin{array}{c}n_{i-1},\\ a_{i-1},\end{array}\! \begin{array}{c}0,\\ a_{i-1}, \end{array}\! \begin{array}{c}1, \\ a_{i+1},\end{array}\! \cdots; t\right)\delta_{n_i,1}\delta_{n_{i+1},0}\delta_{a_{i},a_{i+1}}-P(\{n_i, a_{i}\}; t)\delta_{n_i,1}\delta_{n_{i+1},0}\delta_{a_{i},a_{i+1}}\Big]~,
\end{split}
\eeq
and

\beq
L_R(P)=-\lambda\sum_i\delta_{n_{i},1}\delta_{n_{i+1},1}P(\{n_i, a_{i}\}; t)+ \lambda \sum_i (L^{\tau\tau\tau}_i + L^{\tau\trivial\tau}_i + L^{\trivial\tau\trivial}_i + L^{\tau\tau\trivial}_i +L^{\trivial\tau\tau}_i)~,
\eeq
where $D$ is the diffusion constant (with the assumption that the lattice constant is one), $\lambda$ is the reaction rate, and
\beq \notag
\begin{split}
L_i^{\tau\tau\tau}&=P\left(\cdots,\! \begin{array}{c}n_{i-1},\\ \tau,\end{array}\! \begin{array}{c}1,\\ \tau, \end{array}\! \begin{array}{c}1, \\ \tau,\end{array}\! \cdots; t\right)[\varphi^{-1}\delta_{n_i,0}\delta_{n_{i+1},0}+\frac{1}{2}\varphi^{-2}(\delta_{n_i,0}\delta_{n_{i+1},1}+\delta_{n_i,1}\delta_{n_{i+1},0})]\delta_{a_{i-1},\tau}\delta_{a_{i},\tau}\delta_{a_{i+1},\tau} \\
L_i^{\tau\trivial\tau}&=P\left(\cdots,\! \begin{array}{c}n_{i-1},\\ \tau,\end{array}\! \begin{array}{c}1,\\ \trivial, \end{array}\! \begin{array}{c}1, \\ \tau,\end{array}\! \cdots; t\right)[\varphi^{-2}\delta_{n_i,0}\delta_{n_{i+1},0}+\frac{1}{2}\varphi^{-1}(\delta_{n_i,0}\delta_{n_{i+1},1}+\delta_{n_i,1}\delta_{n_{i+1},0})]\delta_{a_{i-1},\tau}\delta_{a_{i},\tau}\delta_{a_{i+1},\tau} \\
L_i^{\trivial\tau\trivial}&=P\left(\cdots,\! \begin{array}{c}n_{i-1},\\ \trivial,\end{array}\! \begin{array}{c}1,\\ \tau, \end{array}\! \begin{array}{c}1, \\ \trivial,\end{array}\! \cdots; t\right)\delta_{n_i,0}\delta_{n_{i+1},0}\delta_{a_{i-1},\trivial}\delta_{a_{i},\trivial}\delta_{a_{i+1},\trivial} \\
L_i^{\tau\tau\trivial}&=P\left(\cdots,\! \begin{array}{c}n_{i-1},\\ \tau,\end{array}\! \begin{array}{c}1,\\ \tau, \end{array}\! \begin{array}{c}1, \\ \trivial,\end{array}\! \cdots; t\right)\frac{1}{2}(\delta_{n_i,0}\delta_{n_{i+1},1}\delta_{a_{i-1},\tau}\delta_{a_{i},\tau}\delta_{a_{i+1},\trivial}+\delta_{n_i,1}\delta_{n_{i+1},0}\delta_{a_{i-1},\tau}\delta_{a_{i},\trivial}\delta_{a_{i+1},\trivial}) \\
L_i^{\trivial\tau\tau}&=P\left(\cdots,\! \begin{array}{c}n_{i-1},\\ \trivial,\end{array}\! \begin{array}{c}1,\\ \tau, \end{array}\! \begin{array}{c}1, \\ \tau,\end{array}\! \cdots; t\right)\frac{1}{2}(\delta_{n_i,0}\delta_{n_{i+1},1}\delta_{a_{i-1},\trivial}\delta_{a_{i},\trivial}\delta_{a_{i+1},\tau}+\delta_{n_i,1}\delta_{n_{i+1},0}\delta_{a_{i-1},\trivial}\delta_{a_{i},\tau}\delta_{a_{i+1},\tau})~.
\end{split}
\eeq
\end{widetext}

To gain more understanding from this master equation, let us first consider a simple case where the initial fusion tree is the all-$\tau$ state, \ie $\{a_{i}=\tau\}$ at $t=0$. 
It is easy to see that $\{a_{i}=\tau\}$ for all $t>0$, so it is sufficient to characterize the system by using only the position DOF, \ie anyon occupation number $\{ n_i\}$. In other words, now $P(\{n_i, a_i=\tau\}; t)=P_2^{{\rm all}-\tau}\cdot P_1(\{n_i\}; t)$, where $P_2^{{\rm all}-\tau}$ is the all-$\tau$ distribution of the fusion tree and is time independent, and $P_1(\{n_1\}; t)$ describes the probability distribution of the position DOF and has nontrivial dynamics. In terms of $P_1$, the master equation reduces to
\begin{widetext}
\beq \label{eq: reaction-diffusion position all-tau}
\begin{split}
    \partial_t P_1(\{n_{i}\}; t)
    =&D\sum_i\big[P_1(\cdots,\!1_i,\! 0_{i+1},\! \cdots; t)\delta_{n_{i}, 0}\delta_{n_{i+1}, 1}+P_1(\cdots,\! 0_i,\! 1_{i+1},\! \cdots; t)\delta_{n_{i}, 1}\delta_{n_{i+1}, 0}\\ 
    &\qquad\ \ 
    -P_1(\{n_{i}\}; t)(\delta_{n_{i}, 1}\delta_{n_{i-1},0}+\delta_{n_{i},0}\delta_{n_{i+1},1})\big]\\
    +&\lambda\sum_i\big[P_1(\cdots, 1_i, 1_{i+1}, \cdots; t)\frac{1}{2}\varphi^{-2}(\delta_{n_{i}, 0}\delta_{n_{i+1}, 1}+\delta_{n_{i}, 1}\delta_{n_{i+1}, 0})\\
 &\qquad\ \ 
 +P_1(\cdots, 1_i, 1_{i+1}, \cdots; t)\varphi^{-1}\delta_{n_{i}, 0}\delta_{n_{i+1}, 0}-\delta_{n_i,1}\delta_{n_{i+1},1}P_1(\{n_{i}\}; t)\big]~.
\end{split}
\eeq
\end{widetext}

Interestingly, this is precisely the master equation describing the reaction-diffusion dynamics of a hybrid of $A+A\rightarrow 0$ and $A+A\rightarrow A$ processes of classical particles, where two adjacent particles can annihilate with probability $p_{\AAtoO}=\varphi^{-1}$ (the $A+A \rightarrow 0$ process) or coagulate with probability $p_{\AAtoA}=\varphi^{-2}$ (the $A+A \rightarrow A$ process) if they react. 
The decay rate of the particle density can be solved exactly, given by $\rho(t) = c/\sqrt{8\pi Dt}$, where
\begin{equation}\label{eq: prefactor}
    c = \frac{2}{p_{\AAtoA}+2p_{\AAtoO}}~.
\end{equation}
See Refs.~\cite{Henkel1995, Henkel1996, Henkel2003} for the derivation of the general result, and we review and specialize it to our case in Appendix~\ref{app:similarity}.
This indeed is consistent with the numerical results shown in Fig.~\ref{fig:diffusion2}, where $c=2/r_1$ with $r_1=\varphi^{-2}+2\varphi^{-1}$ for the all-$\tau$ case.

Next, consider a more general case where $a_{i}$ is not necessarily $\tau$ initially. 
Denote $P_1(\{n_{i}\};t)\equiv \sum_{\{a_i\}} P(\{n_{i},a_{i}\};t)$ as the probability of a given set of the anyon occupation number $\{n_{i}\}$, and $P_2(\{a_{i}\}|\{n_{i}\};t)$ as the conditional probability of a given set of the fusion tree configuration $\{a_{i}\}$, when the anyon occupation numbers are given by $\{n_{i}\}$. 
Then $P(\{n_i, a_i\};t)=P_2(\{a_{i}\}|\{n_{i}\};t)\cdot P_1(\{n_{i}\};t)$. 
To simplify the situation, we will consider internally super-steady states where $P_2(\{a_{i}\}|\{n_{i}\};t)=P_2(\{a_{i}\}|\{n_{i}\})$ is time independent. 
Note that the all-$\tau$ state is such an internally super-steady state, so the following analysis applies to it. Besides the all-$\tau$ state, there can in principle be other internally super-steady states. 
For such an internally-steady state, the master equation reduces to
\beq
\partial_tP(\{n_i, a_i\})=P_2(\{a_{i}\}|\{n_{i}\})\cdot\partial_t P_1(\{n_{i}\})~.
\eeq

Below, we consider the completely random distribution as the initial condition for the internal DOF.
Motivated by the discussion in Sec. \ref{subsec:pure-reaction_random}, we conjecture that such a distribution is internally super-steady and therefore for all times,  $P_2(\{a_{i}\}|\{n_{i}\})=1/F_N$, where $N=\sum_i n_i$.
(And $P_2(\{a_{i}\}|\{n_{i}\})=0$ if the fusion tree $\{a_i\}$ is not physical.) Combining these equations with Eq.~\eqref{eq: master equation} and summing over configurations of $\{a_{i}\}$, we obtain the effective master equation for $P_1(\{n_{i}\})$ as
\begin{widetext}
\beq \label{eq: reaction-diffusion position completely random}
\begin{split}
\partial_tP_1(\{n_{i}\}; t)
    =&D\sum_i\big[P_1(\cdots, 1_i, 0_{i+1}, \cdots; t)\delta_{n_{i}, 0}\delta_{n_{i+1}, 1}+P_1(\cdots, 0_i, 1_{i+1}, \cdots; t)\delta_{n_{i}, 1}\delta_{n_{i+1}, 0}\\
    &\qquad\quad -P_1(\{n_{i}\}; t)\delta_{n_{i}, 1}\delta_{n_{i+1},0}+\delta_{n_{i},0}\delta_{n_{i+1}, 1})\big]\\
    +&\lambda\sum_i\big\{P_1(\cdots, 1_i, 1_i, \cdots; t)[(\delta_{n_{i}, 0}\delta_{n_{i+1},1}+\delta_{n_{i}, 1}\delta_{n_{i+1},0})\frac{1}{2}p_{\AAtoA}
    +\delta_{n_{i}, 0}\delta_{n_{i+1}, 0}p_{\AAtoO}]\\
    &\qquad\quad
    -\delta_{n_i,1}\delta_{n_{i+1},1}P_1(\{n_{i}\}; t)\big\}~,
\end{split}
\eeq 
\end{widetext}
where $p_{\AAtoO}=\varphi^{-1}p_{\tau\tau\tau}+\varphi^{-2}p_{\tau\trivial\tau}+p_{\trivial\tau \trivial}$ and $p_{\AAtoA}=\varphi^{-2}p_{\tau\tau\tau}+\varphi^{-1}p_{\tau\trivial\tau}+p_{\tau\tau \trivial}+p_{\trivial\tau\tau}$.
Here $p_{\tau\tau\tau}$ etc are again the distribution of the five length-3 segments, where we ignore sites with $n_i=0$ when counting the number of segments.

This is again precisely the equation describing a classical hybrid reaction-diffusion process, where two adjacent particles annihilate with probability $p_{\AAtoO}$ and coagulate with probability $p_{\AAtoA}$. 
Again, the prefactor of the particle density $\rho(t)=c/\sqrt{8\pi D t}$ at late times can be solved exactly, given by Eq.~(\ref{eq: prefactor}).
For the completely random distributions, the probabilities of the five length-3 segments are given in Appendix~\ref{app: completely random state}, which gives us $p_{\AAtoO}=(\varphi+2)/(4\varphi+3)$ and $p_{\AAtoA}=(3\varphi+1)/(4\varphi+3)$, resulting in $c=2/r_2$, agreeing with the numerical result in Fig.~\ref{fig:diffusion2}.
This also supports our assumption that the completely random state is an internally super-steady fusion tree distribution in the reaction-diffusion process.

From these results, we see a nontrivial interplay between classical and quantum behaviors in the Fibonacci reaction-diffusion dynamics, at least when the initial fusion tree configurations are either all-$\tau$ or completely random. The nonlocal nature of the internal DOF makes this interplay particularly interesting.

It is worth noting that, in principle, for any initial state, one can always sum over the fusion tree configurations in Eq.~(\ref{eq: master equation}) and obtain an effective ``master equation" for $P_1(\{n_i\}; t)$. 
However, the resulting $p_{\AAtoA}$ and $p_{\AAtoO}$ will generally depend on $t$ and $\{n_i \}$. The mapping to the effective classical hybrid dynamics can only work if $p_{\AAtoA}$ and $p_{\AAtoO}$ are independent of $t$ and $\{n_i\}$.

\subsection{Two-point correlation functions}\label{subsec:2pt_corr}

\begin{figure}
    \includegraphics[width=\columnwidth]{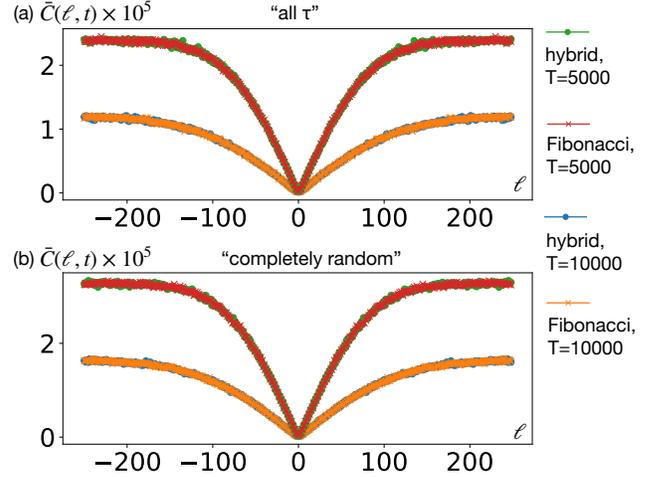}
    \caption{The two point correlation function $\bar{C}(\ell ,t)$ of the Fibonacci reaction-diffusion dynamics with (a) all-$\tau$ initial fusion tree and (b) completely random initial fusion tree, compared with the classical hybrid dynamics with the corresponding parameters. The agreement is remarkable. 
    Note that the data does not contain $\ell = 0$ point.
    }
    \label{fig:2ptcorr}
\end{figure}

Now we examine the two-point correlation functions of the reaction-diffusion dynamics with the all-$\tau$ and completely random initial fusion trees, which are not only interesting to study on their own right, but also provide further nontrivial corroboration of the mapping derived in the previous subsection.
In particular, we consider, for $\ell \neq 0$, 
\begin{equation}
    C(x,\ell,t) = \la n_{x}(t) n_{x+\ell}(t) \ra~,
\end{equation}
where $n_{x}(t)$ is the number of the particles on site $x$ at time $t$ and $\la \hdots \ra$ denotes the average over the quantum trajectories, which are sampled from $1000$ realizations in the numerical simulation.
We further consider the spatial average $\bar{C}(\ell,t)\equiv \frac{1}{|I|}\sum_{x \in I} C(x,\ell,t)$, where $I=(L/4,3L/4]$ and $|I|=L/2$.

The results are shown in Fig. \ref{fig:2ptcorr}. At large $\ell$, we expect $\bar{C}(\ell,t) \approx \rho^2(t)$, which is indeed the case.
At small $\ell$, the dip of the correlation function reflects the fact that the particles have higher chance to be annihilated when they are close in space.

We also compare the results of the Fibonacci dynamics to the classical hybrid $A+A\rightarrow 0$ and $A+A\rightarrow A$ dynamics, with probability $p_{A+A\rightarrow 0}$ for the former and probability $p_{A+A\rightarrow A}$ for the latter. For the Fibonacci dynamics with initial condition given by the all-$\tau$ and completely random fusion tree, we compare it with a hybrid classical dynamics with $(p_{\AAtoO}, p_{\AAtoA})=(\varphi^{-1}, \varphi^{-2})$ and $(p_{\AAtoO}, p_{\AAtoA})=((\varphi+2)/(4\varphi+3), (3\varphi+1)/(4\varphi+3))$, respectively. According to the mapping discussed in the last subsection, the 2-point correlation functions of the Fibonacci dynamics and the hybrid classical dynamics should agree, which is indeed the case, as shown in Fig. \ref{fig:2ptcorr}. This remarkable agreement provides a rather nontrivial check of the mapping.

In passing, we note that the calculated $C(\ell, t)$ in Fig. \ref{fig:2ptcorr} is compatible with a scaling form, $C(\ell, t)=\frac{1}{t}\cdot f(\ell/\sqrt{Dt})$, where $f$ is a universal function and it has been calculated in the context of classical reaction-diffusion dynamics \cite{Masser2001}.

\subsection{Perturbing the initial fusion tree configuration}\label{subsec:perturb_ini_diff}

\begin{figure}
    \includegraphics[width=\columnwidth]{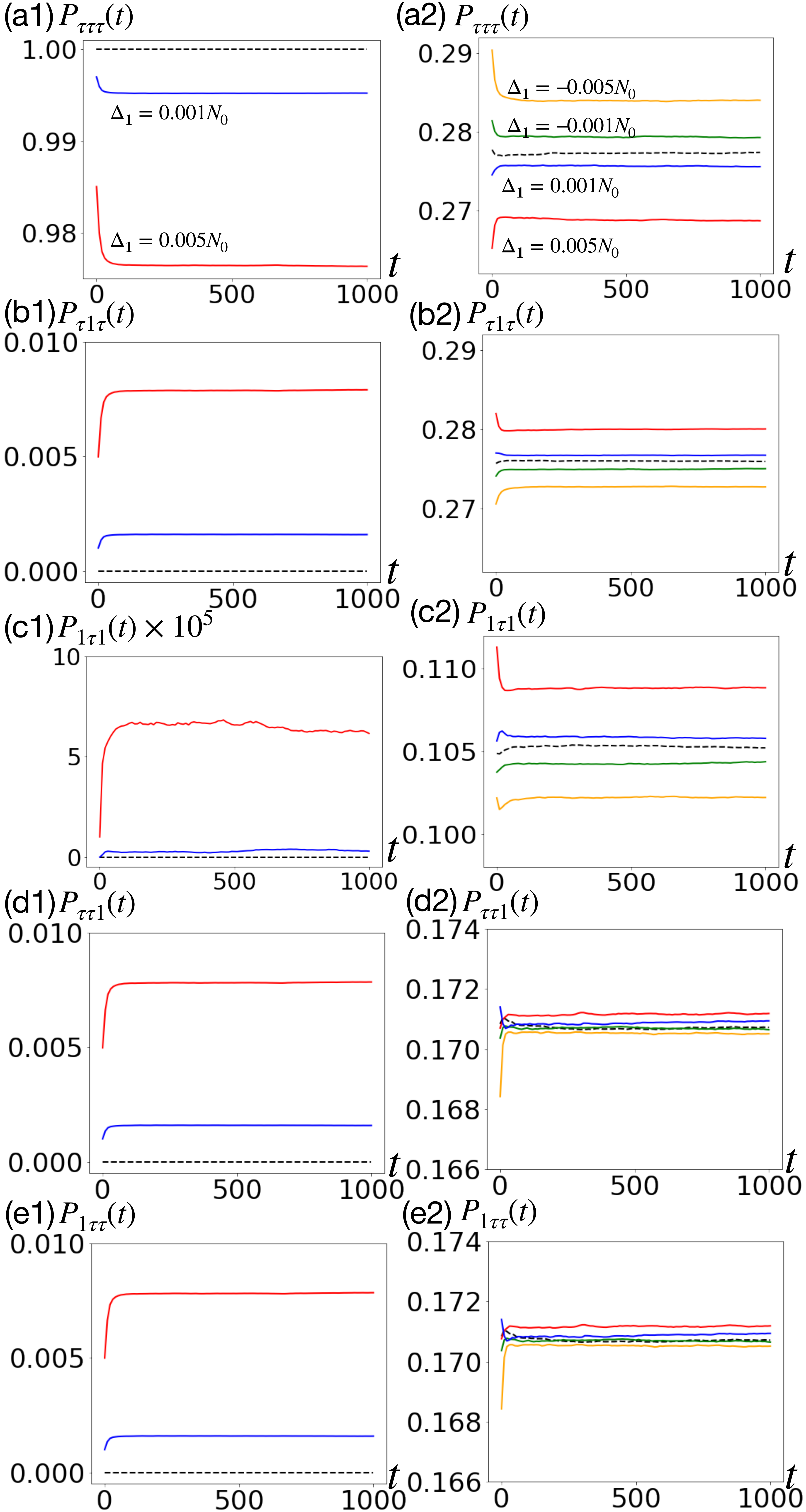}
    \caption{The probability distribution of the length-3 segments with perturbations to the initial fusion tree configurations.
    Left panel (a1)-(e1): perturbed all-$\tau$ fusion tree. Right panel(a2)-(e2): perturbed completely random fusion tree. 
    The black dashed lines denote the results for unperturbed fusion trees. Note the shorter time scale in the figures compared to the previous figures.}
    \label{fig:pert_diff}
\end{figure}

Similar to the pure-reaction dynamics, here we also study the effect of small perturbations to the two initial fusion tree configurations we have discussed.
Again, we consider perturbations of $\Delta_{\trivial}$ in a way consistent with the fusion rules. We randomly pick $\Delta_{\trivial} > 0$ numbers of $\tau$'s replaced with $\trivial$; while pick $|\Delta_{\trivial}|$ of $\trivial$'s replaced with $\tau$ if $\Delta_{\trivial} < 0$.
We examine the probability distributions of the length-3 segments as a function of time shown in Fig.~\ref{fig:pert_diff}.

On the left panel, we show the result of the perturbed all-$\tau$ configurations. We see that the probability distributions flow away from the unperturbed distribution denoted by the black dashed lines, suggesting an instability of the all-$\tau$ fusion tree. 
We notice that the probability distributions approach some other potentially internally-steady distributions in relatively short time scales.

On the right panel, we show the results of the perturbed completely random configurations. 
We again see that the probability distributions flow towards the unperturbed distribution initially, which suggests that the completely random configuration is likely to be stable. 
In addition, we observe that the probability distributions of the length-3 segments approach some other possibly steady values that are different from the ones given by the completely random distribution. This suggests that there could be other stable internally steady states close to the completely random state.
A more careful detailed study in the vicinity of the completely random state is therefore warranted for future work, in order to uncover the precise nature of those potential stable internally steady states and explore the dynamics transitions between them. 

In passing, we note that the change in the probability distributions over time on the right panels are likely due to statistical errors or time fluctuations.

\section{Summary and discussions}\label{sec:discussions}

In this paper, we have studied the reaction-diffusion dynamics of Fibonacci anyons in one dimension. As we have highlighted in Introduction, the Fibonacci anyons are intrinsically quantum and nonlocal, and besides the position DOF, they also have a nonlocal internal DOF. The interplay between the position and internal DOF gives rise to nontrivial interplay between classical and quantum behaviors in the reaction-diffusion dynamics. Our study adds to the growing list of rich nonequilibrium many-body phenomena.

In the first part (Sec.~\ref{sec:pure-reaction}), we study the pure-reaction dynamics without diffusion, where we focus on the number of anyons and the structure of the fusion tree as a function of time.
The instantaneous decay rate of the anyon number is controlled by the fusion tree configuration. 
In particular, the five length-3 probability distributions determines the decay rate via Eqs.~\eqref{eq: decay rate master}, \eqref{eqn:p1} and \eqref{eqn:p2}. 
We find that there are at least two time-independent fusion tree distributions: the all-$\tau$ and completely random fusion trees.
These two fusion tree distributions result in a constant decay rate of the anyon number, given as $r_1$ and $r_2$, via Eqs.~\eqref{eq: decay rate all-tau no diffusion} and \eqref{eq: r2}, respectively. 
We also further study the stability of such fusion tree distributions. 
From the analytical argument based on the rate equations and the numerical results, we conclude that the all-$\tau$ configuration is unstable while the completely random is likely to be stable.

With the intuition for the dynamics of the internal DOF gained in Sec.~\ref{sec:pure-reaction}, in the second part (Sec.~\ref{sec:reaction-diffusion}), we study the reaction-diffusion dynamics, and we mainly focus on the cases where the initial fusion tree is either all-$\tau$ or completely random. Again, we are interested in the density of anyons and the structure of the fusion tree as a function of time. 
We indeed observe the generic $\rho(t) = c/\sqrt{8\pi Dt}$ behavior at late times for the Fibonacci anyons, as expected for any 1d reaction-diffusion dynamics. 
From the numerical simulation, we confirm that $P_R$ is irrelevant in the prefactor $c$ and $D$, while $P_D$ is irrelevant in the prefactor $c$. 
We therefore establish again the superuniversality of the $1/\sqrt{Dt}$ behavior in the Fibonacci reaction-diffusion dynamics and the universality of the prefactor $c$. We stress that universal late-time dynamics associated with the all-$\tau$ and completely random initial fusion trees are distinct, although they correspond to a single steady state of the underlying dynamics, \ie the state with a vanishing anyon density.

We also write down a mater equation describe the reaction-diffusion dynamics in the continuum time limit. For the all-$\tau$ and completely random fusion tree distributions, we can ``integrate out" the internal DOF and obtain an effective hybrid classical reaction-diffusion dynamics for the position DOF. 
Crucially, in deriving such effective dynamics, we have assumed the time independence and translation invariance of the probability distribution of the length-3 segments.
While for any general probability distribution of the fusion tree, one can in principle still integrate out the fusion tree part (internal DOF) and obtain an effective master equation for the position DOF, the effective probabilities of the $A+A\rightarrow A$ and $A+A \rightarrow 0$ dynamics will depend on time or the position DOF in general, which renders this treatment less useful. 
Therefore, the all-$\tau$ and completely random fusion tree distributions are special states of the internal DOF, where the dynamics of the position DOF can be mapped to a hybrid classical reaction-diffusion dynamics.
With this mapping, we determine the prefactors $c$ in $\rho(t)$, via Eq.~\eqref{eq: prefactor}. 
This mapping is further supported numerically by comparing the two-point correlation functions of the Fibonacci dynamics to the corresponding hybrid classical dynamics. This nontrivial interplay between classical and quantum behaviors in the Fibonacci reaction-diffusion dynamics is perhaps one of the most interesting results of this paper, especially because the internal DOF is nonlocal.

Finally, we also examine the stability of the all-$\tau$ and completely random fusion trees against perturbations. 
We observe that, similar to the pure-reaction dynamics, the all-$\tau$ configuration is unstable, while the completely random configuration is likely to be stable. 
We note that the transient time for the fusion tree (internal DOF) is much shorter than the time scale of the position DOF.
It is also worth noting that the time scale in the pure-reaction dynamics and the reaction-diffusion dynamics are different. 
In the pure-reaction dynamics, at each time step, only one pair of anyons go through the reaction. 
On the other hand, in the reaction-diffusion dynamics, there are $O(N(t))$ pairs of anyons going through the reaction after some transient time, where the anyons are starting to diffuse into each other. So the pure-reaction and reaction-diffusion dynamics naturally have different time scales in the simulations.

In conclusion, here we study the reaction-diffusion dynamics of Fibonacci anyons, and obtain some universal results exactly. We find an interesting interplay between classical and quantum behaviors in this dynamics, arising from the interplay between the usual position DOF and the internal DOF. The internal DOF originates from the quantum and nonlocal nature of the Fibonacci anyons, and their non-Abelian character plays an important role in the interplay. For the reaction-diffusion dynamics of Abelian anyons, we expect that it to be qualitatively similar to that of classical domain walls.

Below we briefly discuss some future problems.

In this paper, we have mainly focused on two types of initial fusion tree configurations and their vicinity, and we have assumed a definite initial anyon number in the simulations. The dynamics with the all-$\tau$ initial configuration is relatively easy to understand. For the completely random initial fusion tree, it is helpful to have a more rigorous argument to show that it is an internally steady state. 

It is also interesting to study the case with other initial fusion tree configurations and/or an indefinite initial anyon number, and explore the dynamics transition (as briefly discussed in Introduction) between the different late-time dynamics as the initial condition varies. In particular, we note that the initial conditions we have studied can all be viewed as classical stochastic mixtures of states, and there can in principle be more general quantum superpositions, in terms of the fusion tree and/or anyon number configuration. Considering these more general initial states may bring in new phenomena.

In addition, besides the physical quantities we have studied, one can consider other types of quantities, such as other types of correlation functions of the position DOF and other aspects of the internal DOF, which may also show intriguing universal behaviors. Furthermore, a field theoretic approach to the reaction-diffusion dynamics of the Fibonacci anyons may bring in new insights. Due to the nontrivial internal DOF of these non-Abelian anyons, we expect this field theory to have a different structure compared to the field theories for the classical reaction-diffusion dynamics \cite{Tauber2005}. As other future directions, it is interesting to incorporate other types of reactions (such as creations of anyons) into the reaction-diffusion dynamics, and even generalize it to 2d. Note that although in Sec. \ref{subsec:reaction-diffusion} we have argued that braiding is unimportant for the universal late-time reaction-diffusion dynamics in 1d, it may be important in 2d.

Our framework can be straightforwardly generalized to study the reaction-diffusion dynamics of other types of anyonic systems. As discussed above, the reaction-diffusion dynamics of Abelian anyons is expected to be qualitatively similar to that of classical domain walls, and more interesting behaviors are expected to appear in non-Abelian anyonic systems. On the one hand, Fibonacci anyons are the simplest non-Abelian anyon, in that it has only a single nontrivial anyon. On the other hand, it is also already rather nontrivial, in that they cannot be described by states in a Hilbert space that can be factorized into a tensor product of local Hilbert spaces, so it is tricky to define quantum entanglement in a physically motivated way.

It is interesting to compare our study with the reaction-diffusion dynamics of Majorana defects, as studied in Ref.~\cite{Nahum2019}. The Majorana system studied there involves 3 types of anyons, $\trivial$, $\psi$ and $\sigma$, standing for the trivial, Abelian fermionic and non-Abelian anyons, respectively. The $\sigma$ anyon is the counterpart of the Fibonacci anyon $\tau$ studied in this paper. However, the Majorana system can be described by a Hilbert space that can be factorized into a tensor product of local Hilbert spaces, so in certain sense it is less nonlocal than our Fibonacci system, and quantum entanglement can be defined in the usual way there. Relatedly, the fusion product of two $\sigma$'s can only be $\trivial$ and $\psi$, both of which are Abelian. This is also connected with a difference in the details of the reaction process. In our work, if two adjacent Fibonacci anyon $\tau$'s fuse into a $\tau$, we choose to convert the two original $\tau$'s into a single final $\tau$. This choice has no analog in the Majorana system, because two $\sigma$'s cannot fuse into another single $\sigma$. Instead, there both $\sigma$'s are annihilated (retained) if they fuse into $\trivial$ ($\psi$). One can ask what happens if we also choose to retain the two $\tau$'s if they fuse into $\tau$, which can in principle be designed by hand in a realization of our system in terms of a hybrid quantum circuit, although it may be microscopically unnatural in a realization at the interface between a topological order and the vacuum, because in that case it appears that we need the energy of $\tau$ to be negative, which makes the system tend to retain both $\tau$'s, no matter whether they fuse into $\trivial$ or $\tau$. At any rate, if we do choose to retain (remove) two $\tau$'s if they fuse into $\tau$ ($\trivial$), we expect the late-time dynamics to be different from the ones studied here. For example, from the perspective of the mapping of the dynamics of the position DOF to a hybrid classical dynamics, if such a mapping still works, it is natural to expect the mapped hybrid classical dynamics to contain a {\it pure-diffusion} dynamics and an $A+A\rightarrow 0$ reaction-diffusion dynamics, which appears to be quite different from our setup, where the hybrid classical dynamics contains an $A+A\rightarrow 0$ and $A+A\rightarrow A$ reaction-diffusion dynamics. It is interesting to analyze in detail the Fibonacci dynamics in that setup of the reaction processes, and to understand the Majorana dynamics in Ref.~\cite{Nahum2019} from the angle of this mapping. We leave these problems to future works.

\begin{acknowledgments}

We thank Chong Wang and Timothy Hsieh for helpful discussion. We acknowledge support from Perimeter Institute for Theoretical Physics and Compute Canada (\href{https://www.computecanada.ca}{www.computecanada.ca}).
This research was supported in part by Perimeter Institute for Theoretical Physics. 
Research at Perimeter Institute is supported in part by the Government of Canada through the Department of Innovation, Science and Economic Development Canada and by the Province of Ontario through the Ministry of Colleges and Universities.

\end{acknowledgments}

\onecolumngrid
\appendix

\section{Brief review of Fibonacci anyons} \label{app: review of Fibonacci}

To be self-contained, in this appendix, we briefly review the structure of the Hilbert space of a chain of Fibonacci anyons that will be used in this paper, and we refer the readers to Refs. \cite{Feiguin2006, Nayak2007, 10.1143/PTPS.176.384} for more details.

One choice of the basis states of these Fibonacci anyons can be represented by the left-most fusion tree in Fig. \ref{fig: setup} (a), and it is labeled by the $a$'s in each segment of the tree. Here each vertical line can be thought of as a site that is occupied by a Fibonacci anyon, and each $a_i=\trivial$ or $a_i=\tau$ means the fusion product of $a_{i-1}$ and $\tau$. More concretely, reading the fusion tree from left to right, $a_1$ can be viewed as the fusion result of the $\tau$ in the left-most horizontal segment and the first vertical $\tau$, $a_2$ can be viewed as the fusion product of $a_1$ and the second vertical $\tau$, and so on. The fusion rules relevant to this paper include:
\beq \label{eq: fusion rules}
\trivial\times\trivial=\trivial,
\quad
\trivial\times \tau=\tau,
\quad
\tau\times\tau=\trivial+\tau~.
\eeq
According to these fusion rules, when the left-most horizontal segment is fixed to be $\tau$ and there are $N$ vertical legs, there are in total $F_N\equiv\frac{1}{2\varphi-1}\left[\varphi^{N+1}+\varphi^{N}-(1-\varphi)^{N+1}-(1-\varphi)^{N}\right]$ possible states if the $N$-th horizontal segment is not fixed, $H_N\equiv\frac{1}{2\varphi-1}\left[\varphi^{N+1}-(1-\varphi)^{N+1}\right]$ possible states if the $N$-th horizontal segment is fixed to be $\tau$, and $H_{N-1}$ states if the $N$-th horizontal segment is fixed to be $\trivial$. Note here the left-most horizontal segment is counted as the {\it zeroth} horizontal segment, and we will always fix the zeroth segment to be $\tau$. Also note that these $F$ and $H$ are Fibonacci arrays satisfying $F_{N+2}=F_{N+1}+F_N$ and $H_{N+2}=H_{N+1}+H_N$.

We can also perform basis changes by local $F$-moves, as shown in the middle of Fig. \ref{fig: setup} (b). Each $F$-move is applied to a length-3 segment, where the two external segments are fixed to be $a$ and $c$, and $F^{a\tau\tau}_c$ is a matrix (see Fig. \ref{fig: setup} (b)). For Fibonacci anyons, due to the fusion rules in Eq. \eqref{eq: fusion rules}, it is easy to verify that (i) if $a=c=\trivial$, then $b=b'=\trivial$; (ii) if $a=\trivial$ and $c=\tau$, or $a=\tau$ and $c=\trivial$, then $b=b'=\tau$; (iii) if $a=c=\tau$, $b$ and $b'$ can be either $\trivial$ or $\tau$. In the first two cases, the $F$-matrices can be taken to be the identity one-by-one matrix. In the last case, the relevant $F$-matrix is given by \cite{Feiguin2006}
\beq
[F^{\tau\tau\tau}_\tau]_{\trivial\trivial}=\varphi^{-1},
\quad
[F^{\tau\tau\tau}_\tau]_{\trivial\tau}=\varphi^{-\frac{1}{2}},
\quad
[F^{\tau\tau\tau}_\tau]_{\tau\trivial}=\varphi^{-\frac{1}{2}},
\quad
[F^{\tau\tau\tau}_\tau]_{\tau\tau}=-\varphi^{-1}~.
\eeq
From these $F$-matrices one can deduce the probability of obtaining a fusion outcome for each length-3 segment, as shown in Fig. \ref{fig: length-3 segments}.

\section{Completely random state: an efficient way to generate it and its properties} \label{app: completely random state}

In this appendix, we derive some useful properties of the completely random state. These properties are applied in the main text to justify that the completely random state is an asymptotic state, and they are also used to efficiently generate a completely random state in the numerical calculations.

\subsection{An efficient way to numerical generate the completely random distribution}

First, we discuss an efficient way to numerically generate a state from the completely random ensemble.

As discussed in Appendix \ref{app: review of Fibonacci}, the number of states in a system with $N$ Fibonacci anyons grows exponentially with $N$, so it is inefficient to directly numerically pick up a state with equal probabilities from all such states. Instead, it is more efficient to build up the fusion tree representing a state segment by segment, from left to right. More precisely, for an integer $1\leqslant n\leqslant N$, if the $(n-1)$-th segment is $\trivial$, then the $n$-th segment must be $\tau$. If the $(n-1)$-th segment is $\tau$, then the $n$-th segment is either $\tau$ or $\trivial$, with probabilities $F_{N-n}/F_{N-n+1}$ and $F_{N-n-1}/F_{N-n+1}$, respectively. Note that we always fix the zeroth segment to be $\tau$.

To see that the above procedure will correctly generate a state in the completely random ensemble, all we need is to verify that the probability of obtaining a specific configuration of the fusion tree agrees with that in the completely random ensemble. The probability to get a fusion tree with segments corresponding to $a_1, a_2, \cdots, a_N$ can be written as
\beq
p_{a_1a_2\cdots a_N}=p_{a_1}\cdot p_{a_2|a_1}\cdot p_{a_3|a_1a_2}\cdots p_{a_N|a_1a_2\cdots a_{N-1}}~,
\eeq
where $p_{a_n|a_1a_2\cdots a_{n-1}}$ is the conditional probability that the $n$-th segment is $a_n$, if the previous $n-1$ segments are given by the sequence of $a_1, a_2, \cdots, a_{n-1}$. If $a_{n-1}=\trivial$, by the fusion rules Eq. \eqref{eq: fusion rules}, $a_n$ is necessarily $\tau$. If $a_{n-1}=\tau$, $a_i$ can be either $\tau$ or $\trivial$, which allows in total $F_{N-n}$ and $F_{N-n-1}$ states for all possible choices of $a_{n+1}, a_{n+2}, \cdots, a_N$, respectively. In the completely random ensemble, each state has an equal probability, so the above algorithm gives rise to precisely the correct $p_{a_n|a_1a_2\cdots a_{n-1}}$ for completely random ensemble, and thus also yields a state with the correct probability in this ensemble.

\subsection{Probabilities of a length-$\ell$ segment}

Next, we discuss the probability of a length-$\ell$ segment in the completely random ensemble, assuming that the number of Fibonacci anyons is large, \ie $N\gg 1$.

To begin, consider length-$1$ segments, which can be either $\tau$ or $\trivial$. In this case, we are interested in $p_\tau$ and $p_\trivial$, the probabilities of these two types of length-1 segments. To this end, we first calculate $p_{\tau}(n)$ and $p_\trivial(n)$, the probability for the $n$-th segment to be $\tau$ and $\trivial$, respectively. In the completely random ensemble, it is not hard to see that
\beq
\begin{split}
&p_\tau(n)=\frac{H_nF_{N-n}}{F_N}\\
&p_\trivial(n)=\frac{H_{n-1}F_{N-n-1}}{H_nF_{N-n}+H_{n-1}F_{N-n-1}}~,
\end{split}
\eeq
where $H_nF_{N-n}$ and $H_{n-1}F_{N-n-1}$ are the numbers of states if the $n$-th segment is $\tau$ and $\trivial$, respectively. As a sanity check, one can indeed verify that they add up to the total number of states, \ie $H_nF_{N-n}+H_{n-1}F_{N-n-1}=F_N$ for any $n$. So
\beq
\begin{split}
&p_\tau=\frac{1}{N}\sum_{n=1}^Np_\tau(n)=\frac{\varphi}{2\varphi-1}\\
&p_1=\frac{1}{N}\sum_{n=1}^Np_\tau(n)=\frac{1}{2\varphi-1}\cdot\frac{1}{\varphi}~,
\end{split}
\eeq
where the limit $N\rightarrow\infty$ is taken in the last step.

Next, let us consider a length-$\ell$ segment, with $\ell\geqslant 2$. Such segments can be classified into 4 classes, depending on whether its left-most and right-most segments are $(\tau,\tau)$, $(\tau,\trivial)$, $(\trivial,\tau)$ and $(\trivial,\trivial)$. For these 4 classes, there are in total $H_{\ell-1}$, $H_{\ell-2}$, $H_{\ell-2}$ and $H_{\ell-3}$ states, respectively.

Suppose the left-most segment of a length-$\ell$ segment is the $n$-th segment of the entire system. For a given configuration inside this length-$\ell$ segment, let us count the number of states of the system for different choices of the configurations {\it outside} this length-$\ell$ segment. It turns out that for {\it each} state in the first, second, third and fourth class, there are respectively in total $H_nF_{N-(n+\ell-1)}$, $H_nF_{N-(n+\ell-1)-1}$, $H_{n-1}F_{N-(n+\ell-1)}$ and $H_{n-1}F_{N-(n+\ell-1)-1}$ states for different choices of the segments outside this length-$\ell$ segment. As a sanity check, one can again verify that these numbers add up to the total number of states, \ie $H_{\ell-1}H_nF_{N-(n+\ell-1)}+H_{\ell-2}\left(H_nF_{N-(n+\ell-1)-1}+H_{n-1}F_{N-(n+\ell-1)}\right)+H_{\ell-3}H_{n-1}F_{N-(n+\ell-1)-1}=F_N$, for all $n$ and $\ell$ with $n+\ell\leqslant N+1$. Therefore, in the completely random ensemble, for such a length-$\ell$ segment that begins with the $n$-th segment of the entire system, the probabilities for finding a state in these 4 classes are
\beq \label{eq: prob micro}
\begin{split}
&p_{\tau, \tau}(n)=\frac{H_nF_{N-(n+\ell-1)}}{F_N}\\
&p_{
\tau,\trivial}(n)=\frac{H_nF_{N-(n+\ell-1)-1}}{F_N}\\
&p_{\trivial, \tau}(n)=\frac{H_{n-1}F_{N-(n+\ell-1)}}{F_N}\\
&p_{\trivial,\trivial}(n)=\frac{H_{n-1}F_{N-(n+\ell-1)-1}}{F_N}~.
\end{split}
\eeq

From Eq. \eqref{eq: prob micro} we find that, in the completely random ensemble, the probability to find a particular length-$\ell$ segment is the same for all such segments within the same class, and for the 4 classes they are given by
\beq \label{eq: prob position averaged}
\begin{split}
    &p_{\tau,\tau}=\frac{1}{N-\ell+1}\sum_{n=1}^{N-\ell+1}p_{\tau,\tau}(n)=\frac{1}{2\varphi-1}\cdot\frac{1}{\varphi^{\ell-2}}\\
    &p_{\tau,\trivial}=\frac{1}{N-\ell+1}\sum_{n=1}^{N-\ell+1}p_{\tau,\trivial}(n)=\frac{1}{2\varphi-1}\cdot\frac{1}{\varphi^{\ell-1}}\\
    &p_{\trivial,\tau}=\frac{1}{N-\ell+1}\sum_{n=1}^{N-\ell+1}p_{\trivial,\tau}(n)=\frac{1}{2\varphi-1}\cdot\frac{1}{\varphi^{\ell-1}}\\
    &p_{\trivial,\trivial}=\frac{1}{N-\ell+1}\sum_{n=1}^{N-\ell+1}p_{\trivial,\trivial}(n)=\frac{1}{2\varphi-1}\cdot\frac{1}{\varphi^{\ell}}
\end{split}
\eeq
Again, in the last step, the limit $N\rightarrow\infty$ is taken, while $\ell$ is fixed.

It is also interesting to note that the completely random state is ``self-averaging in space", in the sense that the probability to find a given segment at any position in the interior of the Fibonacci chain is the same as its average over the positions of this segment, given by Eq. \eqref{eq: prob position averaged}). More precisely, taking the limit $n\rightarrow\infty$ and $N-(n+\ell)\rightarrow\infty$ in Eq. \eqref{eq: prob micro}, which physically corresponds to considering a length-$\ell$ segment in the interior of a long Fibonacci chain, Eq. \eqref{eq: prob micro} becomes
\beq
\begin{split}
&\lim_{n\rightarrow\infty, N-(n+\ell)\rightarrow\infty}p_{\tau, \tau}(n)=\frac{H_nF_{N-(n+\ell-1)}}{F_N}=\frac{1}{2\varphi-1}\cdot\frac{1}{\varphi^{\ell-2}}\\
&\lim_{n\rightarrow\infty, N-(n+\ell)\rightarrow\infty}p_{
\tau,\trivial}(n)=\frac{H_nF_{N-(n+\ell-1)-1}}{F_N}=\frac{1}{2\varphi-1}\cdot\frac{1}{\varphi^{\ell-1}}\\
&\lim_{n\rightarrow\infty, N-(n+\ell)\rightarrow\infty}p_{\trivial, \tau}(n)=\frac{H_{n-1}F_{N-(n+\ell-1)}}{F_N}=\frac{1}{2\varphi-1}\cdot\frac{1}{\varphi^{\ell-1}}\\
&\lim_{n\rightarrow\infty, N-(n+\ell)\rightarrow\infty}p_{\trivial,\trivial}(n)=\frac{H_{n-1}F_{N-(n+\ell-1)-1}}{F_N}=\frac{1}{2\varphi-1}\cdot\frac{1}{\varphi^{\ell}}~,
\end{split}
\eeq
which are identical to the position-averaged probabilities in Eq. \eqref{eq: prob position averaged}. Notice that in deriving Eq. \eqref{eq: prob position averaged}, the length-$\ell$ segments are not assumed to be in the interior of the chain, \ie they can be near the boundaries.

\section{Review of the similarity transformation}\label{app:similarity}
In this appendix we review the similarity transformation that relates the $A+A\rightarrow 0$ dynamics to a hybrid of $A+A\rightarrow 0$ and $A+A\rightarrow A$ dynamics \cite{Krebs1995, Henkel1995, Simon1995, Henkel1996, Henkel2003}.

Consider the master equation that describes a hybrid of $A+A\rightarrow 0$ and $A+A\rightarrow A$ dynamics:
\beq
\frac{\partial P_s}{\partial t}=-\sum_{s'}H_{ss'}P_{s'}~,
\eeq
where the matrix $H$ is defined such that $-H_{ss'}$ is the rate to go from $s'$ to $s$ if $s'\neq s$ (denoted as $T_{s's}$ in the main text), and $H_{ss}=-\sum_{s'\neq s}H_{ss'}$. In this form, the probability distribution $P_s$ is viewed as a quantum wave function in the basis labeled by $s$, \ie $P_s=\la s|P\ra$ with $|P\ra$ the state vector corresponding to $P_s$, and the master equation can be viewed as a Schr\"{o}dinger equation of the quantum wave function $P_s$ under an imaginary time evolution of a non-Hermitian Hamiltonian $H$. An observable represented by an operator $O$ can be written as
\beq
\overline{O}=\la s_0|O|P\ra~,
\eeq
where $\la s_0|=\sum_s\la s|$. Notice that the normalization of these states is determined by requiring $\sum_sP_s=1$.

Suppose the classical particles move on a one dimensional lattice, and each site can host at most one particle. Suppose that only the nearest-neighbor diffusion and reaction are considered, which is expected to be sufficient to study the universal late-time dynamics. Then the Hamiltonian $H=\sum_{i}H_{i,i+1}$, with $H_{i,i+1}$ acting on sites $i$ and $i+1$:
\beq
H_{i, i+1}=D
\left(
\begin{array}{cccc}
0 & 0 & 0 &-2\alpha\\
0 & 1 & -1 & -\gamma\\
0 & -1 & 1 & -\gamma\\
0 & 0 & 0 & 2(\alpha+\gamma)
\end{array}
\right)~,
\eeq
where $D$ is the diffusion constant, $2\alpha D$ is the annihilation rate for $A+A\rightarrow 0$, and $D\gamma$ is the coagulation rate for both $A+A\rightarrow A+0$ and $A+A\rightarrow 0+A$, \ie we consider the case of unbiased coagulation. In writing down this matrix representation of $H_{i,i+1}$, the four basis states are respectively $(0, 0)$ (both sites $i$ and $i+1$ empty), $(A, 0)$ (site $i$ occupied and site $i+1$ empty), $(0, A)$ (site $i$ empty and site $i+1$ occupied) and $(A, A)$ (both sites $i$ and $i+1$ occupied). In this model, the case with $\gamma=0$ and $\alpha\neq 0$ represents the $A+A\rightarrow 0$ dynamics, the case with $\gamma\neq 0$ and $\alpha=0$ represents the $A+A\rightarrow A$ dynamics, and the case with $\gamma\neq 0$ and $\alpha\neq 0$ represents a hybrid of $A+A\rightarrow 0$ and $A+A\rightarrow A$ dynamics, where the ratio of the annihilation rate and coagulation rate is $\alpha/\gamma$.

We will be interested in the case with an uncorrelated uniform initial state, corresponding to a quantum state $|P(t=0)\ra=\otimes_{i=1}^L\left(\begin{array}{c}1-\rho\\\rho\end{array}\right)_i$, written in a basis labeled by the occupation number at each site. For example, $\left(\begin{array}{c}1-\rho\\\rho\end{array}\right)_i$ means that the probability for site $i$ to be empty and occupied is $1-\rho$ and $\rho$, respectively. We will discuss the evolution of the density at a given site, and correlation functions of the densities at different sites. In this basis, the density at site $i$ is represented by
\beq
n_i=
\left(
\begin{array}{cc}
   0 & 0 \\
   0 & 1
\end{array}
\right)~.
\eeq
Clearly, for the initial state considered above, the density is uniform and given by $\rho$.

Notice that under the following similarity transformation:
\beq
\begin{split}
&|P(t=0)\ra\rightarrow |\tilde P(t=0)\ra\equiv B|P(t=0)\ra,\\
&H\rightarrow \tilde H\equiv BHB^{-1},\\
&O\rightarrow \tilde O\equiv OB^{-1},
\end{split}
\eeq
the physical observable is invariant:
\beq
\overline{O}(t)=\la s_0|O|P(t)\ra=\la s_0|Oe^{-Ht}|P(t=0)\ra\rightarrow \la s_0|\tilde Oe^{-\tilde Ht}|\tilde P(t=0)\ra=\overline{O}(t)~.
\eeq
Therefore, this similarity transformation relates different dynamics, characterized by different $H$'s and $|P(t=0)\ra$'s.

In our particular case, if we take $B=\otimes_{i=1}^LB_i$ with
\beq
B_i=
\left(
\begin{array}{cc}
1 & \frac{\gamma}{2(\alpha+\gamma)} \\
0 & \frac{2\alpha+\gamma}{2(\alpha+\gamma)}
\end{array}
\right)~,
\eeq
then $|\tilde P(t=0)\ra=B|P(t=0)\ra=\otimes_{i=1}^L\left(\begin{array}{c}1-\frac{2\alpha+\gamma}{2(\alpha+\gamma)}\rho \\ \frac{2\alpha+\gamma}{2(\alpha+\gamma)}\rho\end{array}\right)_i$, $\tilde H=\sum_i\tilde H_{i,i+1}$ with
\beq
\tilde H_{i,i+1}=D
\left(
\begin{array}{cccc}
 0 & 0 & 0 & -2 (\alpha +\gamma ) \\
 0 & 1 & -1 & 0 \\
 0 & -1 & 1 & 0 \\
 0 & 0 & 0 & 2 (\alpha +\gamma ) \\
\end{array}
\right)
\eeq
and the density operator at site $i$ becomes $\tilde n_i=\frac{2(\alpha+\gamma)}{2\alpha+\gamma}\left(\begin{array}{cc} 0 & 0 \\ 0 & 1\end{array}\right)$. That is to say, up to a factor $\left(\frac{2(\alpha+\gamma)}{2\alpha+\gamma}\right)^m$, an $m$-point correlation function of the densities in the hybrid dynamics under our consideration is identical to that in an $A+A\rightarrow 0$ dynamics with annihilation rate $2D(\alpha+\gamma)$ and initial density $\frac{2\alpha+\gamma}{2(\alpha+\gamma)}\rho$:
\beq
\overline{n_{i_1}n_{i_2}\cdots n_{i_m}}=\left(\frac{2(\alpha+\gamma)}{2\alpha+\gamma}\right)^mC^{(m)}_{i_1i_2\cdots i_m}\left(D, 2D(\alpha+\gamma), \frac{2\alpha+\gamma}{2(\alpha+\gamma)}\rho\right)~,
\eeq
where $C^{(m)}_{i_1i_2\cdots i_m}(D, 2D\alpha, \rho)$ is an $m$-point correlation function of the density in a $A+A\rightarrow 0$ dynamics with diffusion constant $D$, annihilation rate $2D\alpha$ and initial density $\rho$.

This relation is useful in simplifying the analysis. As a simple application, consider the evolution of the density at late times. The above relation indicates that $n(t)=\frac{2(\alpha+\gamma)}{2\alpha+\gamma}\cdot\frac{(Dt)^{-1/2}}{\sqrt{8\pi}}$, where we have used the known results of the late-time density for the $A+A\rightarrow 0$ dynamics: $C_i^{(1)}(D, 2D\alpha, \rho)=\frac{(Dt)^{-1/2}}{\sqrt{8\pi}}$ \cite{Bramson1980, Torney1983, Lushnikov1987}. In Sec. \ref{subsec:master_equation}, after reducing the full dynamics of the Fibonacci anyons to the dynamics of their position DOF in Eqs. \eqref{eq: reaction-diffusion position all-tau} and \eqref{eq: reaction-diffusion position completely random}, we have $\frac{\alpha}{\gamma}=\frac{p_{A+A\rightarrow 0}}{p_{A+A\rightarrow A}}$. Using $p_{A+A\rightarrow 0}+p_{A+A\rightarrow A}=1$, we get $c=\frac{2(\alpha+\gamma)}{2\alpha+\gamma}=\frac{2}{p_{A+A\rightarrow A}+2p_{A+A\rightarrow 0}}$, as given by Eq. \eqref{eq: prefactor}.

\twocolumngrid
\bibliography{ref.bib}


\end{document}